\documentclass[]{IEEE-Con-Sys-mag}
\DeclareMathVersion{sans}
\SetSymbolFont{operators}{sans}{LY1}{\sfdefault}{m}{n}
\SetSymbolFont{letters}{sans}{OML}{zplm}{m}{it}
\SetSymbolFont{symbols}{sans}{OMS}{zplm}{m}{n}
\SetSymbolFont{largesymbols}{sans}{OMX}{zplm}{m}{n}

\usepackage{etex}

\usepackage[]{hyperref}
\usepackage{amsthm}
\usepackage{amssymb}
\let\labelindent\relax
\usepackage{enumitem}
\usepackage{mathtools}
\usepackage{xparse} 

\usepackage[title]{appendix}
\usepackage{environ}
\usepackage{lettrine}

\usepackage{graphicx} 
\usepackage{float}
\usepackage{pgfplots}
\usetikzlibrary{arrows.meta,positioning,calc}
\usepackage{subcaption}
\usepackage{color}
\usepackage{wasysym}

\usepackage{booktabs}  
\usepackage{colortbl}  
\usepackage{tabularx}  
\usepackage{multirow}  
\usepackage[skins]{tcolorbox}  
\usepackage[figuresright]{rotating}  

\usepackage{xcolor}
\definecolor{AppleWhite}{RGB}{255,255,255}
\definecolor{ApplePrimaryCoolGray}{RGB}{116,128,139}
\definecolor{AppleCoolGray1}{RGB}{199,209,214}
\definecolor{AppleCoolGray2}{RGB}{147,174,190}
\definecolor{AppleCoolGray3}{RGB}{124,147,160}
\definecolor{AppleCoolGray4}{RGB}{92,102,109}
\definecolor{AppleCoolGray5}{RGB}{78,93,100}
\definecolor{AppleCoolGray6}{RGB}{53,60,65}
\definecolor{AppleBlack}{RGB}{0,0,0}
\definecolor{AppleSecondaryChartGray}{RGB}{168,168,168}
\definecolor{AppleChartGray2}{RGB}{233,233,233}
\definecolor{AppleChartGray3}{RGB}{211,211,211}
\definecolor{AppleChartGray4}{RGB}{190,190,190}
\definecolor{AppleChartGray5}{RGB}{140,140,140}
\definecolor{AppleChartGray6}{RGB}{102,102,102}
\definecolor{AppleChartGray7}{RGB}{64,64,64}
\definecolor{ApplePrimaryChartBlue}{RGB}{84,151,193}
\definecolor{AppleBlue2}{RGB}{212,229,239}
\definecolor{AppleBlue3}{RGB}{169,202,223}
\definecolor{AppleBlue4}{RGB}{127,177,209}
\definecolor{AppleBlue5}{RGB}{71,130,166}
\definecolor{AppleBlue6}{RGB}{55,99,128}
\definecolor{AppleBlue7}{RGB}{45,72,89}
\definecolor{ApplePrimaryChartGreen}{RGB}{83,172,121}
\definecolor{AppleGreen2}{RGB}{212,234,221}
\definecolor{AppleGreen3}{RGB}{169,213,188}
\definecolor{AppleGreen4}{RGB}{126,193,155}
\definecolor{AppleGreen5}{RGB}{58,140,82}
\definecolor{AppleGreen6}{RGB}{39,102,54}
\definecolor{AppleGreen7}{RGB}{29,58,31}
\definecolor{ApplePrimaryChartYellow}{RGB}{253,195,93}
\definecolor{AppleYellow2}{RGB}{254,240,214}
\definecolor{AppleYellow3}{RGB}{254,224,174}
\definecolor{AppleYellow4}{RGB}{254,210,134}
\definecolor{AppleYellow5}{RGB}{230,168,69}
\definecolor{AppleYellow6}{RGB}{191,131,46}
\definecolor{AppleYellow7}{RGB}{153,107,54}
\definecolor{ApplePrimaryChartOrange}{RGB}{250,151,92}
\definecolor{AppleOrange2}{RGB}{254,229,214}
\definecolor{AppleOrange3}{RGB}{252,203,173}
\definecolor{AppleOrange4}{RGB}{252,178,133}
\definecolor{AppleOrange5}{RGB}{227,121,68}
\definecolor{AppleOrange6}{RGB}{191,87,46}
\definecolor{AppleOrange7}{RGB}{143,59,36}
\definecolor{ApplePrimaryChartRed}{RGB}{227,94,105}
\definecolor{AppleRed2}{RGB}{248,215,217}
\definecolor{AppleRed3}{RGB}{241,174,180}
\definecolor{AppleRed4}{RGB}{234,135,143}
\definecolor{AppleRed5}{RGB}{196,63,77}
\definecolor{AppleRed6}{RGB}{153,35,53}
\definecolor{AppleRed7}{RGB}{102,19,43}
\definecolor{ApplePrimaryChartPurple}{RGB}{161,150,204}
\definecolor{ApplePurple2}{RGB}{231,228,242}
\definecolor{ApplePurple3}{RGB}{208,202,229}
\definecolor{ApplePurple4}{RGB}{185,176,217}
\definecolor{ApplePurple5}{RGB}{128,113,171}
\definecolor{ApplePurple6}{RGB}{89,76,128}
\definecolor{ApplePurple7}{RGB}{62,46,101}
\definecolor{AppleCoolGray}{RGB}{116,128,139}
\definecolor{AppleChartGray}{RGB}{168,168,168}
\definecolor{AppleBlue}{RGB}{84,151,193}
\definecolor{AppleGreen}{RGB}{83,172,121}
\definecolor{AppleYellow}{RGB}{253,195,93}
\definecolor{AppleOrange}{RGB}{250,151,92}
\definecolor{AppleRed}{RGB}{227,94,105}
\definecolor{ApplePurple}{RGB}{161,150,204}

\newcommand{\one}[1]{\textcolor{AppleGreen6}{\textbf{#1}}}  
\newcommand{\two}[1]{\textcolor{ApplePurple5}{\textbf{#1}}}
\newcommand{\three}[1]{\textcolor{AppleOrange5}{\textbf{#1}}}
\newcommand{\four}[1]{\textcolor{AppleBlue5}{\textbf{#1}}}
\newcommand{\five}[1]{\textcolor{AppleRed5}{\textbf{#1}}}

\newcommand{\onem}[1]{\textcolor{AppleGreen6}{#1}}
\newcommand{\twom}[1]{\textcolor{ApplePurple5}{#1}}
\newcommand{\threem}[1]{\textcolor{AppleOrange5}{#1}}
\newcommand{\fourm}[1]{\textcolor{AppleBlue5}{#1}}
\newcommand{\fivem}[1]{\textcolor{AppleRed5}{#1}}

\newenvironment{sidebarcol}[2][framc]{%
  \bgroup
  \renewcommand{\theequation}{\sffamily S\arabic{equation}}%
  
  \let\sectionfont\sdbsectionfont
  \let\subsectionfont\sdbsubsectionfont
  
  \begin{figure}[H]%
  \begin{mdframed}[style=MyFrame, backgroundcolor=#1]%
  \sffamily\fontsize{8pt}{12pt}\selectfont\parindent12pt%
  \bgroup\bfseries\fontseries{bc}\fontsize{12pt}{12pt}\selectfont\let\it\itshape#2\par\vspace*{4pt}\egroup
  \mathversion{sans}%
}{%
  \end{mdframed}\end{figure}\egroup
}

\newenvironment{sidebarfig}[1][framc]{%
  \bgroup
  \begin{figure}[H]%
  \begin{mdframed}[style=MyFrame, backgroundcolor=#1]%
  \centering
}{%
  \end{mdframed}\end{figure}\egroup
}

\newcounter{sidewaystablenum}
\NewDocumentEnvironment{sidewaystableplaceholder}{m}{%
  \begin{sidewaystable*}
  \refstepcounter{sidewaystablenum}%
  \begin{mdframed}[style=MyFrame, backgroundcolor=framc, linewidth=1pt, linecolor=AppleCoolGray4, outerlinewidth=0pt]%
  \raggedright
  \sffamily\fontsize{8pt}{12pt}\selectfont\parindent0pt%
  {\bfseries\fontseries{bc}\fontsize{11pt}{13pt}\selectfont TABLE~\arabic{sidewaystablenum}}%
  \hspace{0.6em}\textit{#1}\par\vspace{6pt}%
  \centering
}{%
  \end{mdframed}
  \end{sidewaystable*}
}

\usepackage{cuted}
\newenvironment{sidebarstrip}[2][framc]{%
  \bgroup
  \renewcommand{\theequation}{\sffamily S\arabic{equation}}%
  
  \let\sectionfont\sdbsectionfont
  \let\subsectionfont\sdbsubsectionfont
  
  \setlength{\multicolsep}{0pt}%
  \begin{strip}%
  \begin{mdframed}[style=MyFrame, backgroundcolor=#1]%
  \sffamily\fontsize{8pt}{12pt}\selectfont\parindent12pt%
  \bgroup\bfseries\fontseries{bc}\fontsize{12pt}{12pt}\selectfont\let\it\itshape#2\par\vspace*{-6pt}\egroup
  \begin{multicols}{2}\mathversion{sans}%
}{%
  \end{multicols}\end{mdframed}\end{strip}\egroup
}

\usepackage{xcolor}

\newif\ifshowedits
\showeditstrue      

\ifshowedits
  
\else
  
\fi

\pubyear{2026}

\begin{document}

\title{Controllable AI\stitle{Generative Models as Dynamical Systems that we can control}}

\author{Carmen Amo Alonso\textsuperscript{1}, Marco Pavone\textsuperscript{1,2}}
\affil{\textsuperscript{1}AeroAstro Department, Stanford University, CA, USA; \textsuperscript{2}NVIDIA, Santa Clara, CA, USA}

\maketitle

 \begin{summary}


 \vspace{3mm}
 
{D}espite the remarkable capabilities of modern generative AI systems, we lack a theoretical foundation for their design, and our ability to control or guarantee their behavior remains limited. Closing these gaps would make AI development more resource-efficient and enable its use in critical infrastructure without excessive overhead. Inspired by the design principles and behavioral guarantees that control and dynamical systems theory have brought to other engineering disciplines, we propose to apply similar ideas to AI. This article develops the idea of \emph{Controllable AI}: a control-theoretic framework in which the generative model is treated as a hierarchy of nested dynamical systems across four nested axes. We show that a generative model decomposes into four dynamical systems each with its own state and transition map, and each admitting its own control-theoretic tools for its design and control. At each level we develop the corresponding dynamical description and connect it to existing techniques at both training and inference time. An accompanying open-source implementation builds every level of the hierarchy as a runnable component. By the end, the reader will be equipped both to deploy Controllable AI techniques on their own generative models and to extend the framework to new architectures and application domains.
\end{summary}

\vspace{5mm}

\chapterinitial{A}rtificial intelligence (AI) has emerged as one of the most transformative engineering innovations of our era. Just as powered flight reshaped transportation in the twentieth century, generative models are reshaping our ability to create, reason, and interact with information in the twenty-first. From accelerating scientific discovery to driving our cars, generative AI is on track to become a foundational layer of societal infrastructure, much as commercial aviation, electrification, and the internet did before it. But like every earlier engineering revolution, this one will benefit from design principles and a mathematical theory of behavior in order to be trusted at the scale of our world infrastructures. Addressing this gap is one of the most important challenges in the field of AI nowadays.

The failure record is extensive and quantitative. A 2024 study of leading legal AI tools found hallucination rates of 17--33\% even for products explicitly marketed as ``hallucination-free''~\cite{magesh2024}, and a broader assessment of general-purpose language models found rates of 58--88\% on legal queries~\cite{dahl2024}; in the same year, a Canadian tribunal held Air Canada liable for a bereavement-fare policy that its customer-service chatbot had misstated~\cite{aircanada}, and clinical documentation studies found that when medical language models hallucinate, roughly 44\% of those hallucinations are classified as \emph{major}---the kind of error that can injure a patient~\cite{clinical2024}. The AI Incident Database has now catalogued over 1,200 reported incidents of AI-related harm or near-harm~\cite{aiid}. At the frontier of the field, the concern has escalated from product quality to institutional control: in June 2026, the U.S.\ Department of Commerce ordered a major AI laboratory to disable two flagship models after evidence of potential jailbreak~\cite{commerce2026fable}. In parallel, cross-laboratory alignment evaluations have documented cases of \emph{agentic misalignment}, i.e., models that take actions their operators did not sanction during long simulated interactions~\cite{anthropicopenai2025}. 
Whether or not one accepts the strongest framings of these risks, the underlying observation is technical and hard to dispute: modern generative models can act in ways their designers did not intend and cannot prevent.  This article lays the groundwork for a  theory of principled AI architectures whose behavior we can design in a principled way and control using tools from control theory. 
\vspace{3mm}

The most prominent approach to align the model's behavior with human preferences relies on modifying the dynamics by changing its parameters (weights) via retraining (often with reinforcement learning from human feedback (RLHF)~\cite{instructgpt2022}). Despite its empirical success, this approach requires substantial compute and the resulting alignment is not robust to small weight modifications~\cite{wei2024brittleness, qi2024finetuning}. Beyond this, the deeper issue is that safety and behavioral requirements in real applications depend on runtime context (the user, the task, the physical environment, etc.) that no training procedure can enumerate in advance. Weight adjustment cannot condition on runtime context that was never enumerated, so it cannot by itself certify context-dependent behavior. Reliable behavior under specification requires a mechanism that acts \emph{when the model runs}, senses what it is doing, and corrects it in real time. That mechanism is feedback control. Importantly, weight updates and runtime corrections are
both feedback; they differ in which state they act on and how fast that state moves. The weight update is the slowest in the hierarchy.

This predicament is not unprecedented. In the mid-twentieth century, aerospace systems, power grids, and chemical plants faced a structurally similar bottleneck: how to guarantee the safety and predictability of high-dimensional, nonlinear, imperfectly modeled systems. The solution was the formalization of \emph{control and dynamical systems theory}. By modeling processes as a time evolution of relevant properties and closing the loop with structured feedback, control engineering established the mathematical foundations that make modern automation safe, efficient, and certifiable~\cite{kalman1960, zames1966, bertsekas2005}. The core insight was not that engineers could fully characterize a jet engine or a chemical reactor analytically, but that they could \emph{sense a small number of behaviorally relevant quantities online and correct them through feedback}, without ever solving the full forward problem. The idea that a system's behavior can be governed without being fully understood is what turns raw capability into a reliable engineering artifact. Furthermore, control theory has already touched machine learning at several important points. The backpropagation algorithm, first formulated as an optimal control method by Kelley in 1960~\cite{kelley1960} before being rediscovered for neural networks decades later, is a direct descendant of the Pontryagin maximum principle. Neural ordinary differential equations~\cite{chen2018neuralode} recast deep networks as continuous-time dynamical systems, borrowing the adjoint method of optimal control directly from control. Model predictive control has been embedded into learned robotic pipelines as a differentiable layer~\cite{amos2018,agrawal2019}. Each of these is a point contact between the two fields. A central gap is a systematic treatment of the generative model itself as a controllable dynamical system. This is the gap we intend to tackle with this article.

We call this treatment \emph{Controllable AI}: the deliberate design and operation of generative models as dynamical systems that can be governed through principled theory and structured feedback, applied not only to the model’s weights but, critically, to the states that the model traverses as it runs. Informally, Controllable AI is AI that \emph{reliably exhibits specified acceptable behaviors and consistently avoids undesirable ones}. Importantly, this can be enforced without retraining, by a feedback controller that acts on the model during generation. Modern generative models are not, however, plants of the kind classical control was built for. They are massive (billions to trillions of parameters), strongly nonlinear, and, critically, \emph{multi-scale}: they are hierarchies of nested dynamical processes operating on separated time scales. Governing this multi-scale complexity requires a corresponding hierarchy of controllers, each acting at the appropriate scale and each providing guarantees of the appropriate kind. Formalizing this hierarchy in terms of dynamical systems is the central technical contribution of this article. We explain how existing active research programs fit into this approach, and discuss the potential of an integrated control framework to provide principled design and control strategies to AI systems. 

The remainder of this article is organized as follows. Section \emph{Generative Models as Dynamical Systems} introduces the five-component hierarchy spanning four axes and casts each dynamic level as a discrete-time dynamical system, establishing the common language and notation used throughout. Section \emph{Dynamics and Control at Training Time} and Section \emph{Dynamics and Control at Inference Time} describe the dynamics and control processes occurring at training time and inference time, respectively. We close by discussing the challenges and opportunities in assembling these controllers into a unified AI architecture that we can control, i.e., \emph{Controllable AI}.



\begin{sidebar}{Why viewing AI as a dynamical system matters}
\label{sidebar:paradigm_shift}

Viewing generative AI as a dynamical system is more than mere re-labeling. It provides a mathematical framework to address two important vulnerabilities of AI: the lack of a principled design and the absence of behavioral guarantees. 

Traditional engineering has long relied on first-principles models to guarantee safety, stability, and predictability in complex systems like aerospace or power grids. AI design, by contrast, remains largely an empirical, ad hoc craft driven by costly trial-and-error benchmarks. By treating AI architectures as dynamical systems, we can begin to apply rigorous control-theoretic tools---such as reachability and optimal control---to understand \textit{why} and \textit{how} intelligent behavior emerges.

\end{sidebar}

\section{Generative Model as a Dynamical System}

Generative models are the development responsible for the boom in AI. In this section, we look at generative models through the lens of dynamical systems. Our claim is that existing architectures are hierarchies of dynamical systems, and training and inference are processes that evolve a state over time. Making this explicit is what the rest of the article builds on. To this end, this section answers four questions, one per subsection and in the order below:

\begin{itemize}
    \item What is a generative model, as a mathematical object?
    \item What architecture implements one, and why is it a hierarchy of dynamical systems?
    \item How is such a hierarchy trained, and what dynamics does training itself follow?
    \item How is it run at inference at different time-scales?
\end{itemize}

\subsection{What are Generative Models?}
\label{sec:what_are_gm}

From proving a theorem \cite{alphaproof2024} to folding laundry \cite{black2024pi0}, the capabilities of generative models are unprecedented. Yet, at their core, generative models embody an idea that control engineers know well: form a one-step-ahead prediction of a signal together with a representation of its uncertainty, and advance it by conditioning on what has been observed so far. Generative models extend this idea to high-dimensional, richly structured data such as text, images, and audio, and apply it recursively.
Generative models operate by learning--through offline data exposure (\emph{training time})--the structure of an entire data distribution in a domain, and are then queried during deployment (\emph{inference time}) to synthesize new instances of that distribution.

\vspace{5mm}

\begin{center}
\fbox{
\parbox{0.9\linewidth}{
\textbf{Definition.}
Let $\mathbf{x} \in \mathcal{X}$ denote a data point drawn from an unknown distribution $p_{\mathrm{data}}(\mathbf{x})$. A \emph{generative model} is a parametric family of distributions $\{p_\theta(\mathbf{x})\}_{\theta}$, where $\theta$ are learned parameters, trained such that $p_\theta \approx p_{\mathrm{data}}$. Once trained, the model can synthesize new data by drawing samples $\mathbf{x} \sim p_\theta(\mathbf{x})$, or, given an auxiliary input $\mathbf{c}$, by sampling from the conditional distribution $p_\theta(\mathbf{x}\mid\mathbf{c})$.
}
}
\end{center}

\vspace{5mm}

Crucially, the definition places no constraint on \emph{how} $p_\theta$ is represented or how samples are drawn from it: it may be an explicit normalized density, an implicit distribution defined only through a sampling procedure, etc. This gives rise to the question of \emph{how does one build an architecture that learns and implements a generative model?} As we see next, most successful architectural choices for generative models lead to hierarchical dynamical systems, where dynamical processes occur along different time axes.

\vspace{-4mm}
\subsection{Architecture: Building a Generative Model}

Having established that a generative model is a parametric distribution $p_\theta \approx p_{\mathrm{data}}$, the immediate question is: how does one build a system that represents such a distribution? While in principle any sufficiently expressive family of functions could serve as $p_\theta$, the high-dimensional and richly structured nature of real-world data makes this a nontrivial representational challenge in practice. The solution is to implement $p_\theta$ not as a monolithic function but as a hierarchical architecture of dynamical systems.

\begin{figure}[H]
    \centering
    \includegraphics[width=0.5\textwidth]{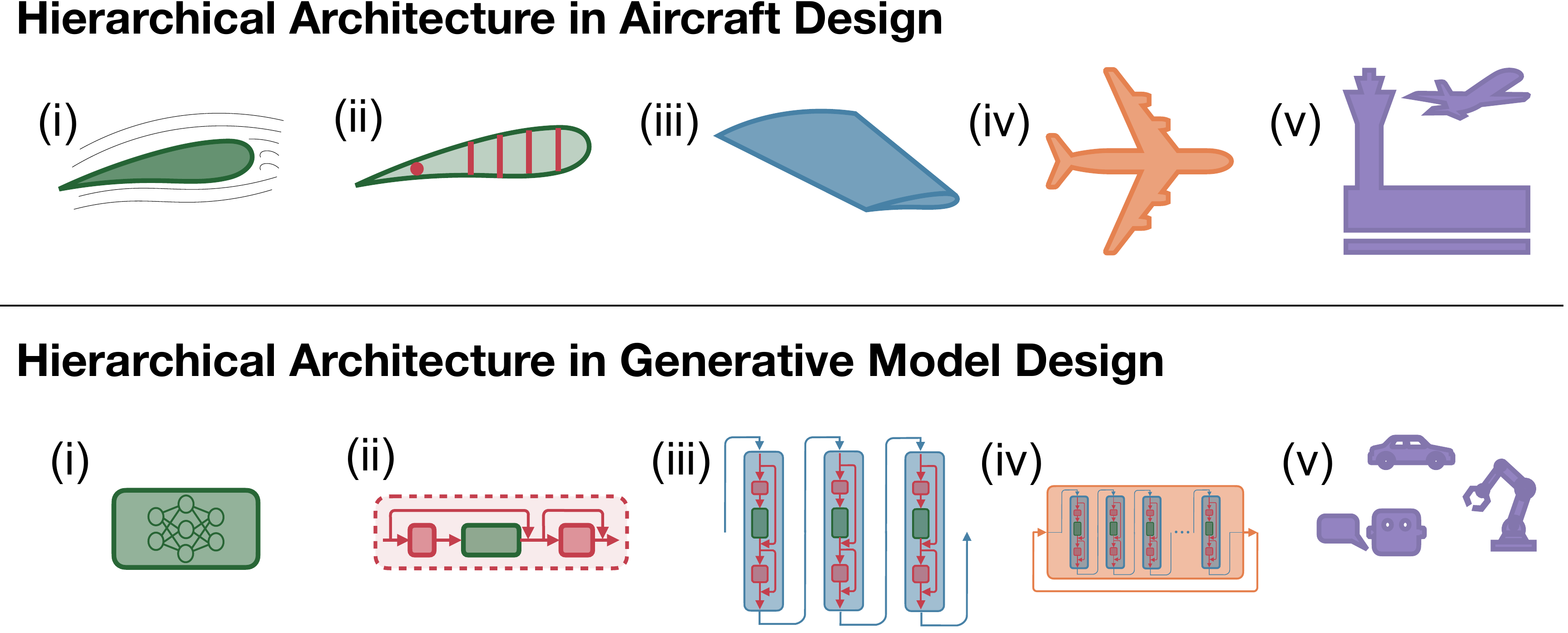}
    \caption{\footnotesize{The hierarchical assembly and flight dynamics of an aircraft, as an analogy to the hierarchical structure of a generative model architecture. At the foundational level, (i) the 2D aerodynamic profile of an airfoil dictates fundamental lift. This is augmented in (ii) with control surfaces, such as flaps, to stabilize and condition the airflow. In (iii), this profile is scaled and repeated spanwise to form a functional 3D wing. These structural components are integrated to build (iv) the fully assembled aircraft, which ultimately engages in (v) the active, closed-loop trajectory of flight. This physical hierarchy is directly analogous to the multi-scale architecture of a generative AI model: (i) represents the core \textbf{token mixer} (e.g., an attention block) that propagates information across the sequence; (ii) adds the \textbf{mixer scaffold} (e.g., LayerNorm and skip connections) that stabilizes the numerical state; (iii) constitutes the sequence of sequentially stacked \textbf{layers}; (iv) forms the \textbf{deep architecture} of the model backbone; and (v) represents the macro-level generative execution loop of inference and \textbf{AI-world interactions}.
    }}
    \label{fig:plane-hierarchy}
\end{figure}

\begin{sidebar} {Why a naive neural network is insufficient} \label{sidebar-neural_network}
A \emph{neural network} is a parametric function $f_\theta : \mathcal{X} \to \mathcal{Y}$ built by composing simple, differentiable operations. In its most basic form, a \emph{fully-connected} (or \emph{dense}) network applies alternating affine maps and pointwise nonlinearities:
\begin{equation}
  \mathbf{h}^{(\ell)} \;=\; \sigma\!\bigl(W^{(\ell)}\,\mathbf{h}^{(\ell-1)} + 
  \mathbf{b}^{(\ell)}\bigr), \quad \ell = 1, \ldots, L,
  \label{eq:mlp_layer}
\end{equation}
where $W^{(\ell)}$ and $\mathbf{b}^{(\ell)}$ are learned weight matrices and biases, and $\sigma$ is a nonlinearity (e.g., ReLU). By the universal approximation theorem, networks of sufficient width can approximate any continuous function on a compact domain to arbitrary accuracy. However, despite their theoretical expressiveness, generic fully-connected networks face a fundamental difficulty when applied to structured, high-dimensional data. 

Consider modeling the joint distribution over a sequence of $T$ tokens, each drawn from a vocabulary of size $V$: the number of parameters required to represent all pairwise interactions grows as $\mathcal{O}(T^2 V^2)$, which is intractable for typical sequence lengths and vocabulary sizes. More critically, a generic network treats every input dimension symmetrically---it cannot exploit the fact that nearby pixels in an image, or nearby tokens in a sentence, are more strongly correlated than distant ones. Without architectural inductive biases that reflect the structure of the data, learning an accurate $p_\theta$ from a finite dataset is statistically and computationally challenging.
\end{sidebar} 

\vspace{-6mm}

\paragraph{\textbf{Time-scale Induced Hierarchy}} 
The hierarchical architecture of dynamical systems that emerges in generative models is far from arbitrary. It arises due to a time-scale separation, where a generation involves dynamics that occur along different axes:

\begin{itemize}
    \item \textbf{Token axis.} When a signal (text, audio, image, etc.) is fed into a generative model, it is often broken down into a sequence, i.e., its $T$-length discrete-time counterpart. Those units, referred to as \emph{tokens}, are often represented as vectors $x_t\in\mathbb R^{d}$, where $t$ indicates their order in the sequence.
    \item \textbf{Depth axis.} Once a signal is fed into a generative model, it is processed sequentially across $L$ steps (implemented as \emph{layers}, as we will discuss later). Each of those steps is captured through \emph{activations}, i.e., matrices $h_\ell\in\mathbb R^{T\times d}$, where $\ell$ is the processing step.
    \item \textbf{Generation axis.} After a signal has gone through the $L$ processing steps, it is often not produced at once but generated over $K$ steps, recurring through the $L$-stack multiple times. Each of those steps yields a partially generated sample (in autoregressive processes) or progressively refined sample (in denoising processes) denoted as the \emph{generation state}, i.e., a matrix $s_k\in\mathbb R^{(T+k)\times d}$, or $s_k\in\mathbb R^{T\times d}$ respectively, where $k$ indicates the generation step.
    \item \textbf{Interaction axis.} Once deployed, a generative model no longer operates on fixed data but \emph{interacts} with the external world in a closed loop over $N$ exchanges. Each of those exchanges is an \emph{interaction}: the model emits an action $a_n$, the environment responds, and the updated context $c_{n+1}$ becomes the next input, where $n$ indicates the interaction step.
\end{itemize}

\paragraph{\textbf{A Hierarchy of Architectural Components}}

At each of the time-scale axes, a dynamical process driven by a specific component emerges. In what follows, we describe the different components that appear at the different levels of granularity, when looking with the lens of dynamical processes at different time scales.

\vspace{.5mm}
\textit{(i) \one{The token mixer.}} Operating at the \textbf{token axis}, the token mixer is the backbone of sequence modeling. Its role is to leverage information from the sequence $x_1,\dots,x_T$ (which can be read as a trajectory) to compute an output. Intuitively, the token mixer propagates information across the sequence: for every position $t$ it produces an output vector $y_t\in\mathbb R^{d}$ built from \emph{all} the tokens it can see, not just from $x_t$. Mathematically, it implements a map $\onem{f} : \mathbb{R}^{T\times d}\!\to\!\mathbb{R}^{T\times d}$ from an input sequence to an output sequence in two steps. First, each token $x_\tau$ is transformed on its own into a \emph{value} $W_V^{\top}x_\tau$, the content it offers. Second, the output at position $t$ is a weighted sum of these values, where the weight on token $\tau$ says how much position $t$ draws from it. Writing this weight as $\onem{[M(}\mathbf{x}\onem{)]_{t\tau}}$, we obtain:

\vspace{-2mm}
\begin{equation}\label{eq:token_mixer}
  \onem{f}(\mathbf{x}) \;=\; \onem{M(}\mathbf{x}\onem{)}\,\mathbf{x}\,\onem{W_V}\,,
  \qquad
  y_t \;=\; \onem{\sum_{\tau=1}^{T} \big[M(}\mathbf{x}\onem{)\big]_{t\tau}\, W_V^{\top}} x_\tau ,
\end{equation}

\begin{sidebarstrip}[AppleChartGray3!45!white]{Example: Axes in a Transformer Language Model}
\label{sidebar:four-axes}

Let us take a dialogue exchange with a language model:

\smallskip
{\scriptsize
\noindent
{\setlength{\fboxsep}{4pt}\fcolorbox{AppleCoolGray3}{white}{%
\color{AppleCoolGray6}\begin{tabular}[t]{@{}l@{}}
\textbf{User:} Book me a flight to Boston on Tuesday.\\
\textbf{Model:} I can do that. Morning or afternoon?\\
\textbf{User:} Morning, and a window seat.\\
\textbf{Model:} ...
\end{tabular}}}%
\hfill
{\setlength{\fboxsep}{4pt}{\begin{tabular}[t]{@{}r@{}}
\emph{input } $c_0$\\ \emph{action } $a_0$\\ \emph{input } $c_1$\\ \emph{action } $a_1$
\end{tabular}}}%
}
\smallskip

Along the \textbf{interaction axis} there are $N=2$ exchanges. Each model line is an action $a_n$; each user line is the world's response, returning as the next input. Notice that each action is not produced at once, but emitted one token at a time:
\vspace{2mm}

\smallskip
{\scriptsize
\newcommand{\tk}[1]{\text{#1}}
\newcommand{\sptk}[1]{\text{\textunderscore #1}}
\newcommand{\slab}[1]{{\color{black}#1}}
\centerline{%
{\setlength{\fboxsep}{4pt}\fcolorbox{AppleCoolGray3}{white}{%
\color{AppleCoolGray6}$\smash[b]{%
\underbrace{\tk{I}}_{\slab{s_1}}\cdot
\underbrace{\sptk{can}}_{\slab{s_2}}\cdot
\underbrace{\sptk{do}}_{\slab{s_3}}\cdot
\underbrace{\sptk{that}}_{\slab{s_4}}\cdot
\underbrace{\tk{.}}_{\slab{s_5}}\cdot
\underbrace{\sptk{Morn}}_{\slab{s_6}}\cdot
\underbrace{\tk{ing}}_{\slab{s_7}}\cdot
\underbrace{\sptk{or}}_{\slab{s_8}}\cdot
\underbrace{\sptk{after}}_{\slab{s_9}}\cdot
\underbrace{\tk{noon}}_{\slab{s_{10}}}\cdot
\underbrace{\tk{?}}_{\slab{s_{11}}}}$}}}

\vspace{16pt}
}

Along the \textbf{generation axis}, each partial reply after step $k$ is the generation state $s_k$. For instance, step 10 was taken with steps 1--9 already fixed, and with nothing after it yet in existence. Notice that each generation step is produced across a series of sequential $L$ steps (implemented as \emph{layers}):

\smallskip
{\scriptsize
\centerline{%
{\setlength{\fboxsep}{4pt}\fcolorbox{AppleCoolGray3}{white}{%
\color{AppleCoolGray6}$h_0 \;\to\; h_1 \;\to\; \cdots \;\to\; h_{L} \;\to\; \text{``ing''}$}}}
}
\smallskip

Along the \textbf{depth axis}, each processing step $\ell$ is captured by an activation $h_\ell$, i.e., a matrix $h_\ell\in\mathbb R^{T\times d}$, where $\ell$ is the processing step. Notice that the activation $h_\ell$ is not a single vector. It is $T$ token vectors side by side,

\smallskip
{\scriptsize
\centerline{%
{\setlength{\fboxsep}{4pt}\fcolorbox{AppleCoolGray3}{white}{%
\color{AppleCoolGray6}$h_\ell \;=\; [\,x_1,\; x_2,\; \dots,\; x_{T}\,], \qquad x_t \in \mathbb{R}^{d}$}}}
}
\smallskip

Along the \textbf{token axis}, each index $t$ indicates every position in the sequence so far. In causal (autoregressive) settings, each token $x_t$ is processed using only the previous tokens up to $t$. The token axis is the innermost and fastest of the four axes.

\end{sidebarstrip}

\noindent where $\mathbf{x} = [x_1,\dots,x_T]^{\top} \in \mathbb{R}^{T \times d}$ is the input sequence of $T$ tokens, $\onem{W_V} \in \mathbb{R}^{d \times d}$ is a learned value (channel) projection applied identically to every token, $\onem{M(}\mathbf{x}\onem{)} \in \mathbb{R}^{T \times T}$ is the token-mixing operator whose entry $\onem{[M(}\mathbf{x}\onem{)]_{t\tau}}$ specifies how much token $x_\tau$ contributes to output position $t$; causal

 \noindent models constrain $\onem{M(}\mathbf{x}\onem{)}$ to be lower-triangular; and $\onem{f(}\mathbf{x}\onem{)} = [y_1,\dots,y_T]^{\top} \in \mathbb{R}^{T \times d}$ is the mixed output sequence.

We note that the map $\onem{f}$ realized by the token mixer appears to be a static one in equation \eqref{eq:token_mixer} because it is expressed in convolutional form. This form best captures most existing token mixing strategies; and as we show in the next sections, very rich dynamics appear when the map is unrolled. We defer this discussion for later, and briefly list different families of token mixers in sidebar \emph{Token Mixer Families}, which mostly differ in how $M(\mathbf{x})$ is parameterized.

\begin{sidebarcol}[AppleGreen6!18!white] {Token Mixer Families} \label{sidebar:token-mixer}

\begin{itemize}
  \item \textbf{Attention \cite{vaswani2017}:} computes pairwise interactions between all tokens via a dense, data-dependent mixing (nonlinear in $\mathbf x$) operator $M(\mathbf{x})$. Because interactions are pairwise, each layer costs $\mathcal{O}(T^2)$ time and memory.
  \item \textbf{Recurrent Networks (RNNs \cite{elman1990rnn}, LSTMs \cite{hochreiter1997lstm}):} process tokens sequentially, maintaining a hidden state  that summarizes past context. They realize the mixing role through this nonlinear recurrence, and the operator $M(\mathbf{x})$ can be seen as the convolution. While they offer better memory efficiency than attention, their sequential nature makes them less parallelizable and thus less computationally efficient.
  \item \textbf{State space models 
   (S6~\cite{gu2023mamba}):} parameterize a linear input-dependent state-space recurrence that corresponds to a structured, lower-triangular matrix $M(\mathbf{x})$. This parameterization combines the parallelizable training of attention (via parallel scan) with the memory-efficient nature of RNNs (via a hidden state). Yet, since information from distant tokens has to be compressed into a fixed-size state vector, they have a bottleneck.
\end{itemize}
\end{sidebarcol} 

\textit{(ii) \five{Mixer scaffold.}} A bare token mixer is unstable during the training process. The mixer scaffold is a set of \emph{per-token} operations that make the architecture trainable. Strictly speaking, they do not introduce dynamics across any of the axes earlier described, and can be seen as a forward map. Abstractly, the scaffold's role is to turn a mixer into a stackable block by composing it in the following \emph{pre-norm} way:
\begin{equation}\label{eq:scaffold-types}
  \fivem{\mathcal{B}}[\,\onem{f}\,] \;=\;
    \big(\fivem{\mathrm{id}} \,\fivem{+}\, \fivem{g}\!\circ\!\fivem{\mathrm{Norm}}\big)
    \,\circ\,
    \big(\fivem{\mathrm{id}} \,\fivem{+}\, \onem{f}\!\circ\!\fivem{\mathrm{Norm}}\big),
\end{equation}
where $\fivem{\mathcal{B}}$ is a functional wrapping the level-(i) token mixer

\begin{sidebarcol}[AppleChartGray3!45!white] {Example: The Token Mixer --- Attention}
\label{sidebar:ex-token-mixer}

Each $x_t$ is an embedded representation of a token, i.e., an element in the sequence. In attention, the output position is a weighted average of the values at earlier positions. The weights are the entries of $M(\mathbf{x})$:

\smallskip
\centering
\begin{tikzpicture}[every node/.style={font=\tiny}]
\newcommand{\cw}{9.2pt}
\fill[white] (0,0) rectangle (11*\cw,-11*\cw);
\foreach \t [count=\ti from 0] in {
 {100,0,0,0,0,0,0,0,0,0,0},
 {35,65,0,0,0,0,0,0,0,0,0},
 {20,25,55,0,0,0,0,0,0,0,0},
 {12,15,25,48,0,0,0,0,0,0,0},
 {10,10,12,30,38,0,0,0,0,0,0},
 {14,8,8,14,18,38,0,0,0,0,0},
 {6,5,5,8,8,60,8,0,0,0,0},
 {8,6,6,8,10,22,10,30,0,0,0},
 {6,5,5,6,8,26,8,14,22,0,0},
 {4,3,3,4,5,14,6,6,80,5,0},
 {5,4,4,6,8,12,6,8,14,20,13}}{
  \foreach \w [count=\ki from 0] in \t {
    \ifnum\w>0
      \fill[AppleGreen6!\w!white] (\ki*\cw,-\ti*\cw) rectangle ++(\cw,-\cw);
      \draw[white,line width=.3pt] (\ki*\cw,-\ti*\cw) rectangle ++(\cw,-\cw);
    \fi}}
\draw[AppleGreen6,line width=.7pt] (0,-9*\cw) rectangle (10*\cw,-10*\cw);
\draw[AppleCoolGray6,line width=.5pt] (0,0) rectangle (11*\cw,-11*\cw);
\node[anchor=west,font=\scriptsize,text=AppleCoolGray6]
  at ({11*\cw+4pt},{-5.5*\cw}) {$=\;M(\mathbf{x})$};
\foreach \i/\lab in {0/{$x_1$ (I)},1/{$x_2$ (\_can)},2/{$x_3$ (\_do)},3/{$x_4$ (\_that)},4/{$x_5$ (.)},5/{$x_6$ (\_Morn)},
                     6/{$x_7$ (ing)},7/{$x_8$ (\_or)},8/{$x_9$ (\_after)},9/{$x_{10}$ (noon)},10/{$x_{11}$ (?)}}{
  \node[anchor=east,text=AppleCoolGray6] at (-2pt,{-\i*\cw-0.5*\cw}) {\lab};
  \node[anchor=west,rotate=90,text=AppleCoolGray6] at ({\i*\cw+0.5*\cw},2pt) {\lab};}
\node[anchor=east,text=AppleGreen6] at (-24pt,{-9*\cw-0.5*\cw}) {$t{=}10$};
\end{tikzpicture}\par

\smallskip
\raggedright
Row $t$ reads across the keys $\tau\le t$. The highlighted row says the
continuation token \textcolor{AppleCoolGray6}{noon} draws almost all of its
value from \textcolor{AppleCoolGray6}{\_after}.
\end{sidebarcol}

\begin{sidebarcol}[AppleChartGray3!45!white] {Example: The Mixer Scaffold}
\label{sidebar:ex-scaffold}
The mixer never acts on the stream directly. It hangs off it as a branch, and
the sum node puts the correction back:
\medskip
\centering
\resizebox{\linewidth}{!}{%
\begin{tikzpicture}[font=\tiny, x=1.12cm,
  nb/.style={draw=AppleRed5,fill=AppleRed5!15!white,rounded corners=1pt,
             minimum height=9pt,minimum width=17pt,inner sep=1pt,font=\tiny},
  mb/.style={draw=AppleGreen6,fill=AppleGreen6!18!white,rounded corners=1pt,
             minimum height=9pt,minimum width=17pt,inner sep=1pt,font=\tiny},
  sm/.style={draw=AppleRed5,fill=white,circle,inner sep=0pt,minimum size=8pt,font=\tiny},
  ar/.style={-{Latex[length=1.1mm]},AppleCoolGray6,line width=.45pt}]
\coordinate (in) at (0,0);
\coordinate (b1) at (0.28,0);
\node[sm] (s1) at (1.95,0) {$+$};
\coordinate (b2) at (2.35,0);
\node[sm] (s2) at (4.05,0) {$+$};
\coordinate (out) at (4.55,0);
\draw[ar] (in) -- (s1); \draw[ar] (s1) -- (s2); \draw[ar] (s2) -- (out);
\node[anchor=east,text=AppleCoolGray6] at (in) {$h_{\ell-1}$};
\node[anchor=west,text=AppleCoolGray6] at (out) {$h_\ell$};
\node[nb] (n1) at (0.90,-.95) {Norm};
\node[mb] (f)  at (1.95,-.95) {$f$};
\node[nb] (n2) at (2.97,-.95) {Norm};
\node[nb] (g)  at (4.05,-.95) {$g$};
\draw[ar] (b1) |- (n1.west);
\draw[ar] (n1) -- (f);
\draw[ar] (f) -- (s1);
\draw[ar] (b2) |- (n2.west);
\draw[ar] (n2) -- (g);
\draw[ar] (g) -- (s2);
\fill[AppleCoolGray6] (b1) circle (.7pt); \fill[AppleCoolGray6] (b2) circle (.7pt);
\node[anchor=south,text=AppleCoolGray6,font=\tiny\itshape] at (3.0,.16) {residual stream (identity path)};
\node[anchor=north,text=AppleGreen6,font=\tiny] at (1.95,-1.32) {token mixer};
\node[anchor=north,text=AppleRed5,font=\tiny] at (4.05,-1.32) {channel MLP};
\end{tikzpicture}}\par
\smallskip
\raggedright
At every one of the $L$ sequential activations, the straight path carries the $T$ token vectors of our example forward unchanged, while $f$ (the attention of the previous example) and the channel MLP $g$ compute what to add to them.
\end{sidebarcol}

 \noindent $\onem{f}$ such that $\fivem{\mathcal{B}}[\onem{f}]:\mathbb{R}^{T\times d}\!\to\! \mathbb{R}^{T\times d}$, $\circ$ indicates composition, and $\fivem{\mathrm{id}} + (\cdot)$ denotes a residual map. Unlike $\onem{f}$, which mixes \emph{across} tokens, all three of the above components are static, i.e., they act \emph{per token}: each is a map $\mathbb{R}^{d}\!\to\!\mathbb{R}^{d}$ applied identically to every token. Hence, the following equalities hold: $[\fivem{\mathrm{Norm}}(\mathbf{z})]_t = \fivem{\mathrm{Norm}}(z_t)$ and $[\fivem{g}(\mathbf{z})]_t = \fivem{g}(z_t)$.

Since the components present in equation \eqref{eq:scaffold-types} that provide a scaffold around the token mixer block are widely used in existing generative model architectures, each has a specific nomenclature and role. We provide a description of these in sidebar \emph{Mixer Scaffold Types}.

\begin{sidebarcol}[AppleRed5!18!white] {Mixer Scaffold Types} \label{sidebar:mixer-scaffold}
\begin{itemize}
  \item \textbf{Skip connection \cite{he2016resnet}:} the identity map $\fivem{\mathrm{id}}:\mathbb{R}^{T\times d}\!\to\!\mathbb{R}^{T\times d}$ composed with the addition operation $\fivem{+}$ provides an additive (residual) \emph{update} to the stream.
  \item \textbf{Normalization \cite{ba2016layernorm,zhang2019rmsnorm}:} the normalization map $\fivem{\mathrm{Norm}}:\mathbb{R}^{d}\!\to\! \mathbb{R}^{d}$ acts on the input to $\onem{f}$ and $\fivem{g}$. Different choices exist for this map, such as LayerNorm ($\mathrm{Norm}(x_t) = \gamma \odot \tfrac{x_t - \mu_t}{\chi_t} + \beta$, with $\mu_t, \chi_t$ the mean and standard deviation across the $d$ features of $x_t$ and $\gamma, \beta \in \mathbb{R}^d$ learned), and RMSNorm ($\mathrm{Norm}(x_t) = \gamma \odot
  x_t / \mathrm{RMS}(x_t)$). 
  \item \textbf{Channel function:} the map $\fivem{g}:\mathbb{R}^{d}\!\to\!\mathbb{R}^{d}$ is a static transform applied independently to each token's vector representation $x_t\in\mathbb{R}^d$. In particular, it recombines information \emph{across the $d$ coordinates} (referred to as channels) of a single token. The standard choice is a \emph{multilayer perceptron} (MLP)~\cite{vaswani2017}, $g(z) \;=\; \sigma(z W_1)\,W_2$, which lifts each token from dimension $d$ to a wider hidden width $d_{\mathrm{MLP}}$ through $W_1\in\mathbb{R}^{d\times d_{\mathrm{MLP}}}$, applies a pointwise nonlinearity $\sigma$ (e.g.\ ReLU or GELU~\cite{hendrycks2016gelu}), and projects back to $d$ through $W_2\in\mathbb{R}^{d_{\mathrm{MLP}}\times d}$. Common variants are the \emph{gated MLP}, e.g.\ SwiGLU~\cite{shazeer2020glu}, $g(z)=(\sigma(zW_1)\odot zW_2)W_3$, in which $\odot$ is the elementwise product and one branch multiplicatively \emph{gates} the other. The channel function may also be absent ($g\equiv 0$) when the architecture folds channel mixing into the mixer itself (e.g.\ Mamba~\cite{gu2023mamba}).
\end{itemize}
\end{sidebarcol}

\textit{(iii) \four{Layers.}} When the token mixer is scaffolded by the mixer scaffold, this becomes \emph{a layer}. Operating at the \textbf{depth axis}, the layer
stack is central to deep learning. Its role is to refine the activations through $L$ successive blocks, $h_0,\dots,h_L$, where $h_\ell\in\mathbb R^{T\times d}$ is the output of the $\ell^{th}$ layer. Intuitively, the layer stack is responsible for propagating information across depth, transforming the input into progressively more abstract representations. Mathematically, it implements a map $\fourm{F} : \mathbb{R}^{T\times d}\!\to\!\mathbb{R}^{T\times d}$ from embedded input to final representation, obtained by composing $L$ blocks:
\begin{equation}\label{eq:layers}
  \fourm{F} \;:=\; \fivem{\mathcal{B}}\big[\onem{f}^{(L)}\big]\circ\cdots
    \circ\fivem{\mathcal{B}}\big[\onem{f}^{(1)}\big],
  \qquad
  h_{l} \;=\; \fivem{\mathcal{B}}\big[\onem{f}^{(\ell)}\big]\big(h_{l-1}\big),
\end{equation}
where $h_{\ell} \in \mathbb{R}^{T\times d}$ is the activation after layer $\ell$, with $h_0$ being the embedded input and $h_L = \fourm{F}(h_0)$ the final representation; $\fivem{\mathcal{B}}[\onem{f}^{(\ell)}]:\mathbb{R}^{T\times d}\!\to\!\mathbb{R}^{T\times d}$ is the level-(ii) block at depth $\ell$, built from its own token mixer $\onem{f}^{(\ell)}$; $L$ is the number of layers (the \emph{depth}); $\circ$ denotes composition; and
$\fourm{F}$ is the resulting composition of stacking $L$ layers. The dynamics across the depth axis are captured in equation~\eqref{eq:layers} via the update rule to the activations $h_\ell$, which gives rise to a time-varying dynamical system with finite horizon $L$.

In contrast to the token axis, where multiple choices of maps $\onem{f}$ are available for the token mixer, in the depth axis fewer choices are possible. The main one is the number of layers $L$, as well as how these layers are organized. We provide a short description in sidebar \emph{Layer Structures}.

\begin{sidebarcol}[AppleBlue5!18!white] {Layer Structures} \label{sidebar:layers}

\begin{itemize}
  \item \textbf{Untied layers~\cite{vaswani2017}:} each map $\onem{f}^{(\ell)}$ is independent from the rest. This means that computing activations through the layers is a collection of sequential steps. 
  \item \textbf{Weight-tied layers~\cite{dehghani2019universal, lan2019albert}:} the same map block $\fivem{\mathcal{B}}[\onem{f}^{(\ell)}]$ is repeated across several layers, turning the stack into a recurrence iterating over $\fivem{\mathcal{B}}[\onem{f}^{(\ell)}]$ $L$ times. 
  \item \textbf{Adaptive depth~\cite{graves2016act,raposo2024mod}:} the number of layers is input-dependent rather than fixed, so $L$ itself becomes context-dependent, i.e., the network allocates more depth to tokens that are harder to process.
\end{itemize}

\end{sidebarcol}

\vspace{-5mm}

\begin{sidebarcol}[AppleChartGray3!45!white] {Example: The Layers}
\label{sidebar:ex-layers}
Each box represents a layer in the architecture, and it is one scaffolded block of the previous example. The arrows illustrate how the state is carried sequentially through the layers:\par
\medskip
{\centering
\resizebox{\linewidth}{!}{%
\begin{tikzpicture}[font=\tiny,
  ly/.style={draw=AppleBlue5,fill=AppleBlue5!25!white,line width=.25pt,
             rounded corners=3pt,minimum width=15pt,minimum height=19pt,inner sep=0pt},
  ar/.style={-{Latex[length=1.0mm]},AppleCoolGray6,line width=.4pt},
  dn/.style={-{Latex[length=0.9mm]},AppleBlue5,line width=.35pt},
  lb/.style={anchor=south,inner sep=1pt,text=AppleCoolGray6}]
\node[ly] (La) at (0.90,0) {};
\node[ly] (Lb) at (2.62,0) {};
\node[ly] (Lc) at (4.60,0) {};
\node[ly] (Ld) at (6.32,0) {};
\node[text=AppleCoolGray6,font=\tiny,inner sep=1pt] (out) at (7.30,0) {\textbf{noon}};
\coordinate (p0) at (0.09,0);
\coordinate (p1) at (1.72,0);
\coordinate (p2) at (3.44,0);
\coordinate (p3) at (5.42,0);
\draw[ar] (p0) -- (La.west);        \coordinate (m0) at ($(p0)!0.5!(La.west)$);
\draw[ar] (La.east) -- (p1);        \coordinate (m1) at ($(La.east)!0.5!(p1)$);
\node[text=AppleCoolGray6] at (1.90,0) {$\cdots$};
\draw[ar] (2.08,0) -- (Lb.west);
\draw[ar] (Lb.east) -- (p2);        \coordinate (m2) at ($(Lb.east)!0.5!(p2)$);
\node[text=AppleCoolGray6] at (3.62,0) {$\cdots$};
\draw[ar] (3.80,0) -- (Lc.west);
\draw[ar] (Lc.east) -- (p3);        \coordinate (m3) at ($(Lc.east)!0.5!(p3)$);
\node[text=AppleCoolGray6] at (5.60,0) {$\cdots$};
\draw[ar] (5.78,0) -- (Ld.west);
\draw[ar] (Ld.east) -- (out.west);  \coordinate (m4) at ($(Ld.east)!0.5!(out.west)$);
\node[lb] at ($(m0)+(0,0.16)$) {$h_0$};
\node[lb] at ($(m1)+(0,0.16)$) {$h_1$};
\node[lb] at ($(m2)+(0,0.16)$) {$h_8$};
\node[lb] at ($(m3)+(0,0.16)$) {$h_{24}$};
\node[anchor=north,text=AppleBlue5,font=\tiny,inner sep=1.5pt] at ($(m4)+(0,-0.36)$) {decode};
\node[text=AppleCoolGray6,anchor=north] at ($(m2)+(0,-0.44)$)
     {\scalebox{1.10}{$h_{8}=[\,\dots\,\text{\_or \_after}\;\textbf{\_the}\,]$}};
\node[text=AppleCoolGray6,anchor=north] at ($(m3)+(0,-0.86)$)
     {\scalebox{1.10}{$h_{24}=[\,\dots\,\text{\_or \_after}\;\textbf{noon}\,]$}};
\draw[dn] ($(m2)+(0,-0.14)$) -- ($(m2)+(0,-0.42)$);
\draw[dn] ($(m3)+(0,-0.14)$) -- ($(m3)+(0,-0.84)$);
\end{tikzpicture}}\par}
\smallskip
Since $h_\ell=[\,x_1,\dots,x_T\,]$, and only the last vector is the one to be predicted, that last row is where the next-token choice lives. At $\ell=8$ the choice is still generic; by $\ell=24$ it has essentially arrived at \textbf{noon}.
\end{sidebarcol}

\textit{(iv) \three{Deep Architecture.}} Operating at the \textbf{generation axis}, the generative process is what makes the stack of layers $\fourm{F}$ a model of the data distribution. Its role is to produce a sample $\mathbf{x}\sim p_\theta$ by repeatedly invoking the stack of layers $\fourm{F}$. This process generates a sequence $s_0,\dots,s_K$ that refines or extends the generation until the sample is complete. Intuitively, generation is responsible for placing the stack of layers in a feedback loop, so that each invocation conditions on what has been produced so far. Mathematically, it implements a map $\threem{\Phi}:s_0\mapsto s_K$ from an initial condition (initial prompt, etc.) $s_0$ to a  sample $s_K=\mathbf x\sim p_\theta$, obtained by iterating the layer stack $K$ times:
\begin{equation}\label{eq:generation}
    \threem{\Phi} = \underbrace{\tilde\phi\circ\cdots\circ\tilde\phi}_{K\ \mathrm{times}},
      \quad \tilde\phi(s) := \threem{\phi}\big(s,\, \fourm{F_\theta}(s)\big), \quad
    s_{k+1} = \tilde\phi(s_k),
\end{equation}
where $k = 0,\dots,K-1$ and $s_k$ is the generated sequence at step $k$ 
; $\threem{\phi}$ is the \emph{decoding rule}, generally stochastic; $s_0\sim p_0$ is the initial draw and $\mathbf{x}=s_K\sim p_\theta$ the generated sequence. The dynamics across the generation axis are captured in equation \eqref{eq:generation} via the update rule for $s_k$, which gives rise to a discrete-time dynamical system with finite horizon $K$.

Different choices of the decoding rule $\threem{\phi}$ give rise to different architecture families. We list the most prominent ones in sidebar \emph{Deep Architecture Structures}.

\begin{sidebarcol}[AppleOrange5!18!white] {Deep Architecture Structures} \label{sidebar:deep-architectures}

\begin{itemize}
  \item \textbf{Autoregressive~\cite{radford2018gpt}}: the decoding rule $\threem{\phi}$ appends one new element to the sequence $s_k$ at each $k$ step, so the sample $s_K$ is built up one token at a time.
  \item \textbf{Masked / iterative decoding~\cite{chang2022maskgit}}: the sequence has fixed length, and $\threem{\phi}$ updates a subset of its positions at each step, completing the sample in a few iterations.
  \item \textbf{Diffusion~\cite{ho2020ddpm}}: starting from pure noise, $\threem{\phi}$ updates the whole fixed-length sequence with a less noisy version at each step, gradually turning noise, $s_0$ into data $s_K$.
  \item \textbf{Flow matching~\cite{lipman2022flow}}: $\threem{\phi}$ advances the sequence by a small deterministic step at each iteration, moving an initial noise sample $s_0$ toward a data sample $s_K$.
\end{itemize}

The degenerate case $K=1$ recovers single-pass generators such as
GANs~\cite{goodfellow2014gan} and VAE decoders~\cite{kingma2014vae}, which
produce a sample in one shot.

\end{sidebarcol}

\vspace{-3mm}

\begin{sidebarcol}[AppleChartGray3!45!white] {Example: Deep Architecture -- Autoregressive}
\label{sidebar:ex-generation}
The layer stack of the previous example runs once per token, and its output is appended to the state fed back in:\par
\medskip
{\centering
\resizebox{\linewidth}{!}{%
\begin{tikzpicture}[font=\tiny,
  st/.style={draw=AppleOrange5,fill=white,rounded corners=2pt,line width=.3pt,
             inner sep=2.5pt,text=AppleCoolGray6},
  bb/.style={draw=AppleOrange5,fill=AppleOrange5!20!white,rounded corners=2pt,
             line width=.3pt,inner sep=2.5pt,text=AppleCoolGray6},
  ar/.style={-{Latex[length=1.0mm]},AppleCoolGray6,line width=.4pt},
  lb/.style={text=AppleOrange5,font=\tiny,inner sep=1.5pt},
  sl/.style={text=black,font=\tiny,inner sep=1.5pt}]
\node[st] (a) at (0,0) {I \_can \_do \_that .};
\node[bb,right=0.62cm of a] (F) {$F_\theta$};
\node[st,right=0.62cm of F] (b) {I \_can \_do \_that . \_Morn};
\draw[ar] (a) -- node[sl,above]{$s_5$} (F);
\draw[ar] (F) -- node[lb,above]{$\phi$} (b);
\coordinate (bd) at ($(b.south)+(0,-0.42)$);
\coordinate (ad) at ($(a.south)+(0,-0.42)$);
\draw[ar] (b.south) -- (bd) -- (ad) -- (a.south);
\node[sl,above] at ($(bd)!0.5!(ad)$) {$s_6$};
\end{tikzpicture}}\par}
\smallskip
Eleven turns around this loop produce the reply. The state grows on every pass, while $\threem{F_\theta}$ is identical each time.
\end{sidebarcol}

\textit{(v) \two{AI--world interactions.}} Operating at the \textbf{interaction axis}, the model is a policy acting in closed loop with the world. Its role is to choose actions (text, images, physical movements, etc.) $a_0,\dots,a_{N-1}$ from what it has observed so far, driving an external environment toward a goal. Intuitively, interaction places the whole generative model in a feedback loop, so that each action it takes changes the context it later observes. Mathematically, it implements the dynamics:
\begin{equation}\label{eq:interaction}
  c_{n+1} \;=\; \twom{\psi}\big(c_n,\ a_n\big),
  \qquad
  a_n = \threem{\Phi}(c_n),
\end{equation}
where $c_n$ is the model's context at step $n$ (collection of all the text sequences/images/movements/etc. so far); $a_n$ is the action produced by one run of the generation map $\threem{\Phi}$; $\twom{\psi}$ is the \emph{interaction rule}, which submits the action to the environment and folds the resulting effects back into the context; and $N$ is the interaction horizon. The dynamics across the interaction axis are captured by equation \eqref{eq:interaction} via the update rule to context $c_n$, which gives rise to a closed-loop dynamical system. In this case, since the loop is closed through the external world, the system is unknown, partially observed, and outside the model's control. 

The interaction with the world leads to vastly different choices and combinations of the architectural components. While doing a review of existing architectures across different domains is outside of the scope of this article, in sidebar \emph{AI Interaction Domains} we list popular examples of how generative models interact with the world.

\begin{sidebarcol}[ApplePurple5!18!white] {AI Interaction Domains} \label{sidebar:interactions}

\begin{itemize}

  \item \textbf{Conversational~\cite{instructgpt2022}}: a generative model and a human interact via text, and possibly images. In our modeling, the environment is a human interlocutor that tries to control the model to produce a given output. Examples include text $\mapsto$ text (Large Language Model), text $\mapsto$ image (text-to-image model), and image $\mapsto$ text (Vision Language Model).
  \item \textbf{Embodied / robotic (VLA)~\cite{brohan2023rt2}}: a generative model produces actions executed via a robot, which interacts with the world. In our modeling, the generative model  controls the environment to achieve a goal.  
  \item \textbf{Autonomous software agents~\cite{schick2023toolformer}}: a generative model produces actions executed against external software (calls to APIs, etc.), which in turn act on databases, services, and other agents. In our modeling, the generative model tries to observe and control this software environment to accomplish a specified task, similar to the robotic example. Often, these agents are equipped with a conversational component as well to include the user in the loop.
\end{itemize}
\end{sidebarcol}

\begin{sidebarcol}[AppleChartGray3!45!white] {Example: Interactions -- Conversation}
\label{sidebar:ex-interaction}
One action is a whole run of the previous loop, and a user
closes the loop by answering:\par
\medskip
{\centering
\resizebox{\linewidth}{!}{%
\begin{tikzpicture}[font=\tiny,
  st/.style={draw=ApplePurple5,fill=white,rounded corners=2pt,line width=.3pt,
             inner sep=2.5pt,text=AppleCoolGray6},
  bb/.style={draw=ApplePurple5,fill=ApplePurple5!20!white,rounded corners=2pt,
             line width=.3pt,inner sep=2.5pt,text=AppleCoolGray6},
  wb/.style={draw=ApplePurple5,fill=ApplePurple5!10!white,rounded corners=2pt,
             line width=.3pt,inner sep=2.5pt,text=AppleCoolGray6},
  ar/.style={-{Latex[length=1.0mm]},AppleCoolGray6,line width=.4pt},
  lb/.style={text=ApplePurple5,font=\tiny,inner sep=1.5pt},
  sl/.style={text=black,font=\tiny,inner sep=1.5pt}]
\node[st] (a) at (0,0) {Book me a flight \dots};
\node[bb,right=0.60cm of a] (P) {$\threem \Phi$};
\node[st,right=0.60cm of P] (b) {I can do that. Morning \dots};
\draw[ar] (a) -- node[sl,above]{$c_0$} (P);
\draw[ar] (P) -- (b);
\node[wb] (W) at ($(a)!0.5!(b) + (0,-0.95)$) {User $\psi$};
\draw[ar] (b.south) |- node[sl,pos=0.86,above]{$c_1$} (W.east);
\draw[ar] (W.west) -| node[sl,pos=0.14,above]{$c_1$} (a.south);
\end{tikzpicture}}\par}
\smallskip

\end{sidebarcol}

\begin{sidewaystableplaceholder}{A summary of the hierarchical architecture of generative models. Each row displays the governing dynamics of one level.}
\label{tab:granularity}

\renewcommand{\arraystretch}{2.3}
\setlength{\tabcolsep}{6pt}
\begin{tabular}{@{} >{\raggedleft\arraybackslash}m{1.3in}
  >{\centering\arraybackslash}m{0.7in}
  >{\raggedright\arraybackslash}m{1.9in}
  >{\centering\arraybackslash}m{1.0in}
  >{\centering\arraybackslash}m{1.4in}
  >{\centering\arraybackslash}m{1.5in} @{}}
\toprule
\multicolumn{1}{c}{\textbf{Component}} & \multicolumn{1}{c}{\textbf{Axis}} & \multicolumn{1}{c}{\textbf{Role}} & \multicolumn{1}{c}{\textbf{State}} & \multicolumn{1}{c}{\textbf{Transition Map}} & \multicolumn{1}{c}{\textbf{Illustration}} \\
\midrule
\cellcolor{AppleGreen6!18!white}{\large\one{Token Mixer}} & \cellcolor{AppleGreen6!18!white}Token &
  \cellcolor{AppleGreen6!18!white}Mixes information \emph{across} tokens $x_1,\dots,x_T$; the backbone every other level is built around. &
  \cellcolor{AppleGreen6!18!white}$x_t\in\mathbb{R}^d$ &
  \cellcolor{AppleGreen6!18!white}\small$\onem{f}:\mathbb{R}^{T\times d}\to\mathbb{R}^{T\times d}$ &
  \cellcolor{AppleGreen6!18!white}\includegraphics[width=\linewidth,height=1.2in,keepaspectratio]{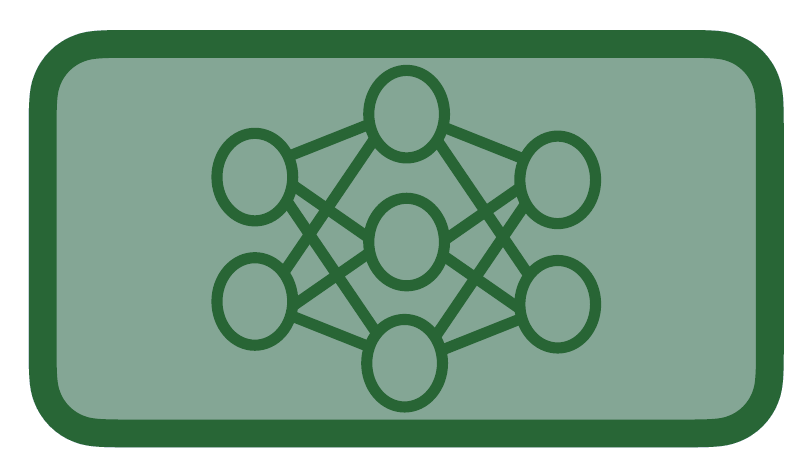} \\
\cellcolor{AppleRed5!18!white}{\large\five{Mixer Scaffold}} & \cellcolor{AppleRed5!18!white}--- (static) &
  \cellcolor{AppleRed5!18!white}Per-token skip connections, normalization, and channel MLP; turns a bare mixer into a trainable, stackable block. Adds no new dynamics. &
  \cellcolor{AppleRed5!18!white}--- &
  \cellcolor{AppleRed5!18!white}\small$\fivem{\mathcal{B}}:\mathbb{R}^{T\times d}\to\mathbb{R}^{T\times d}$ &
  \cellcolor{AppleRed5!18!white}\includegraphics[width=\linewidth,height=1.2in,keepaspectratio]{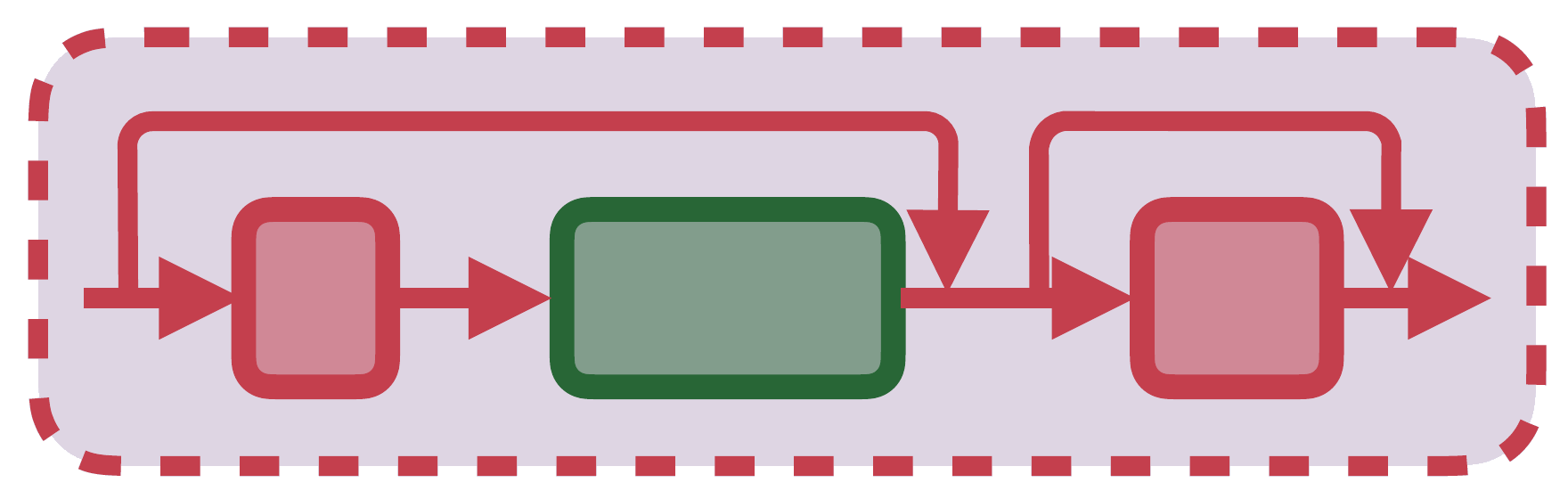} \\
\cellcolor{AppleBlue5!18!white}{\large\four{Layer Stack}} & \cellcolor{AppleBlue5!18!white}Depth &
  \cellcolor{AppleBlue5!18!white}Stacks $L$ scaffolded mixers into the backbone $\fourm{F}$; refines activations $h_0,\dots,h_L$ into progressively abstract representations. &
  \cellcolor{AppleBlue5!18!white}$h_\ell\in\mathbb{R}^{T\times d}$ &
  \cellcolor{AppleBlue5!18!white}\small$\fivem{\mathcal{B}}[\onem{f}^{(\ell)}]:\mathbb{R}^{T\times d}\to\mathbb{R}^{T\times d}$ &
  \cellcolor{AppleBlue5!18!white}\includegraphics[width=\linewidth,height=1.2in,keepaspectratio]{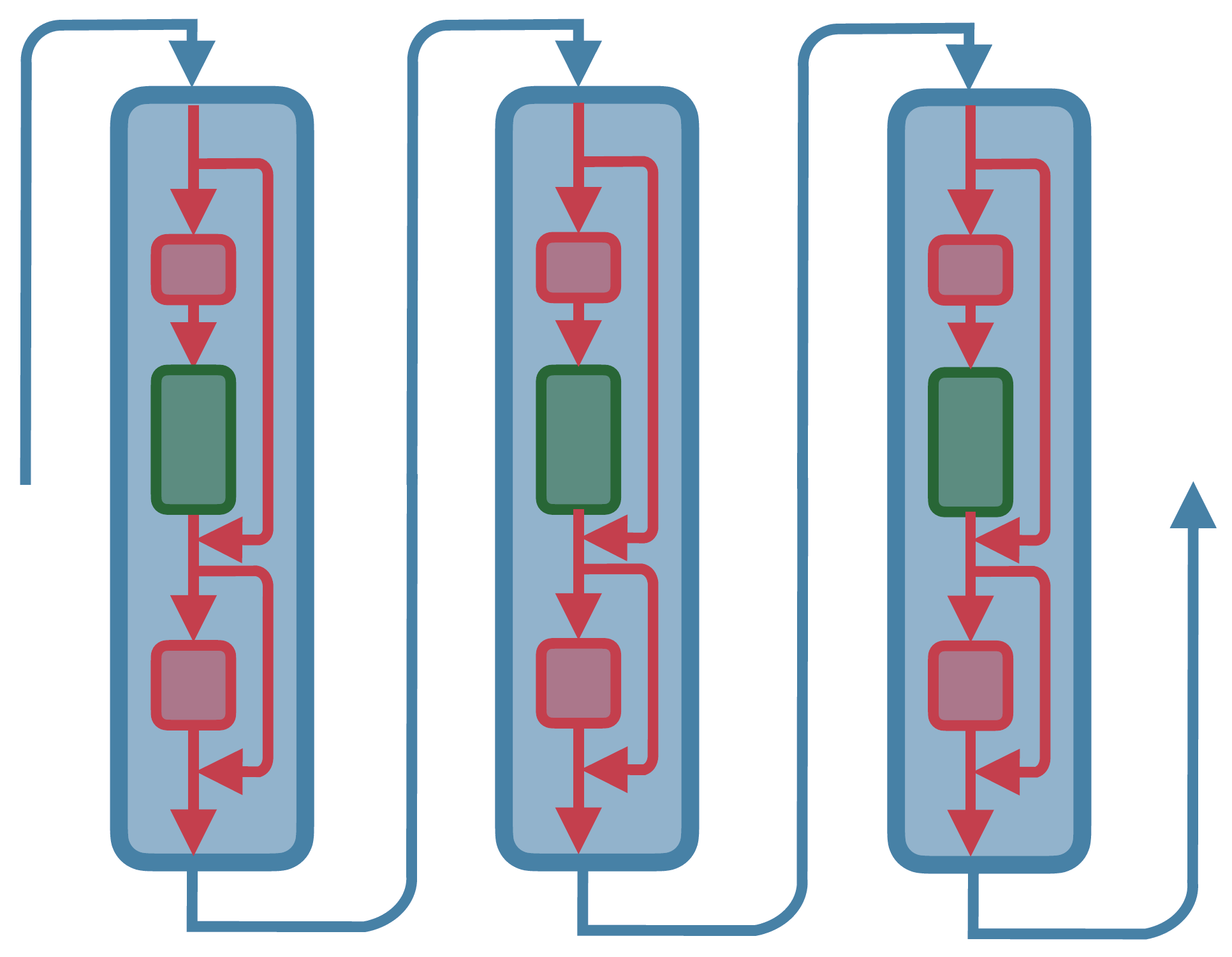} \\
\cellcolor{AppleOrange5!18!white}{\large\three{Deep Architectures}} & \cellcolor{AppleOrange5!18!white}Generation &
  \cellcolor{AppleOrange5!18!white}Iterates the layer stack $K$ times via decoding rule $\threem{\phi}$; turns $\fourm{F}$ into a generative process $\threem{\Phi}$ over the sample. &
  \cellcolor{AppleOrange5!18!white}$s_k\in\mathbb{R}^{T\times d}$ or $s_k\in\mathbb{R}^{(T+k)\times d}$ &
  \cellcolor{AppleOrange5!18!white}\small$\threem{\phi}:\mathbb{R}^{T\times d}\to\mathbb{R}^{T\times d}$ &
  \cellcolor{AppleOrange5!18!white}\includegraphics[width=\linewidth,height=1.2in,keepaspectratio]{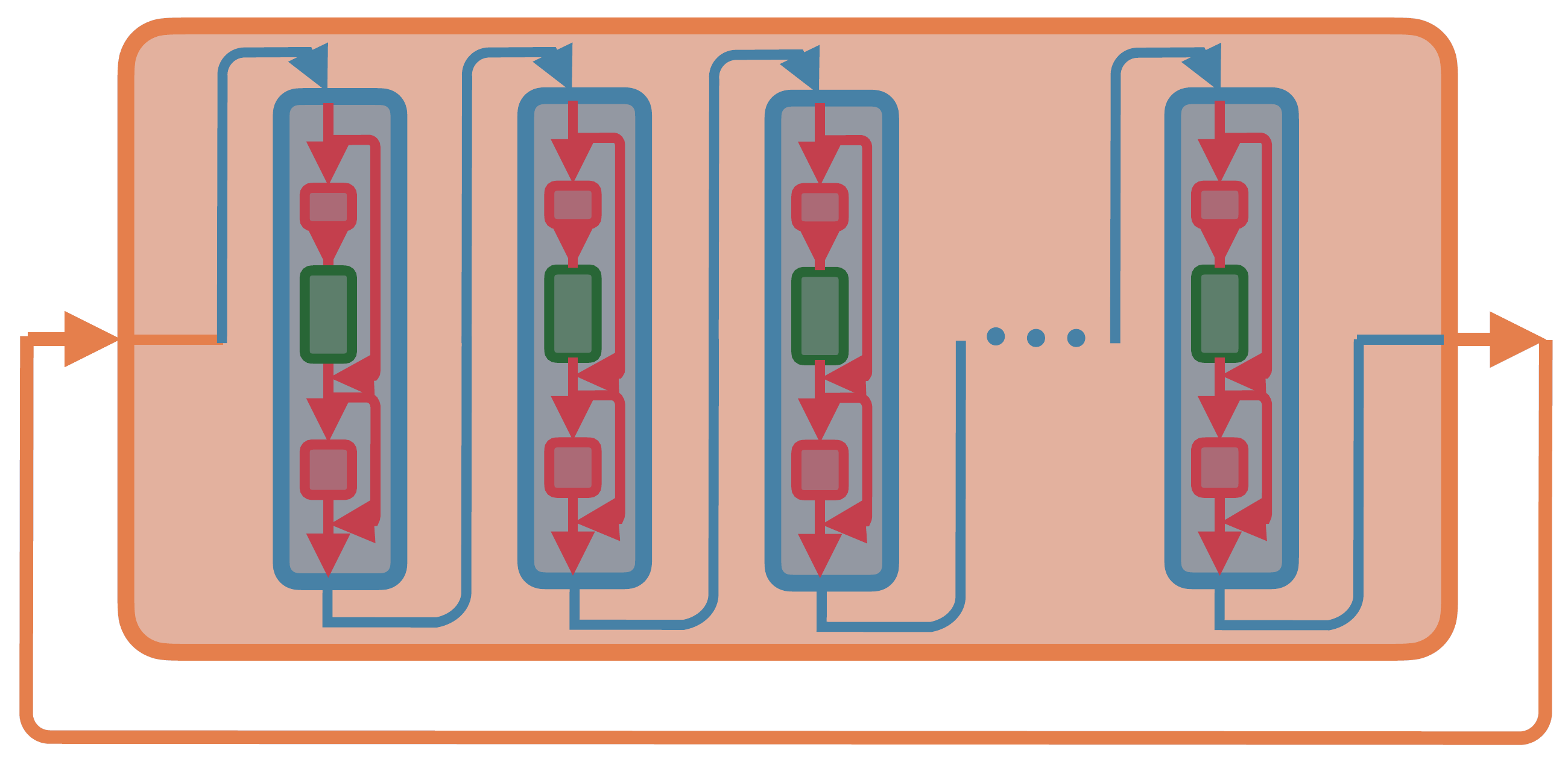} \\
\cellcolor{ApplePurple5!18!white}{\large\two{AI--World Interactions}} & \cellcolor{ApplePurple5!18!white}Interaction &
  \cellcolor{ApplePurple5!18!white}Closes the loop with the external world over $N$ exchanges; each action changes the context the model next observes. &
  \cellcolor{ApplePurple5!18!white}$c_n\in\mathcal{C}$ &
  \cellcolor{ApplePurple5!18!white}\small$\twom{\psi}:\mathcal{C}\times\mathcal{A}\to\mathcal{C}$ &
  \cellcolor{ApplePurple5!18!white}\includegraphics[width=\linewidth,height=1.2in,keepaspectratio]{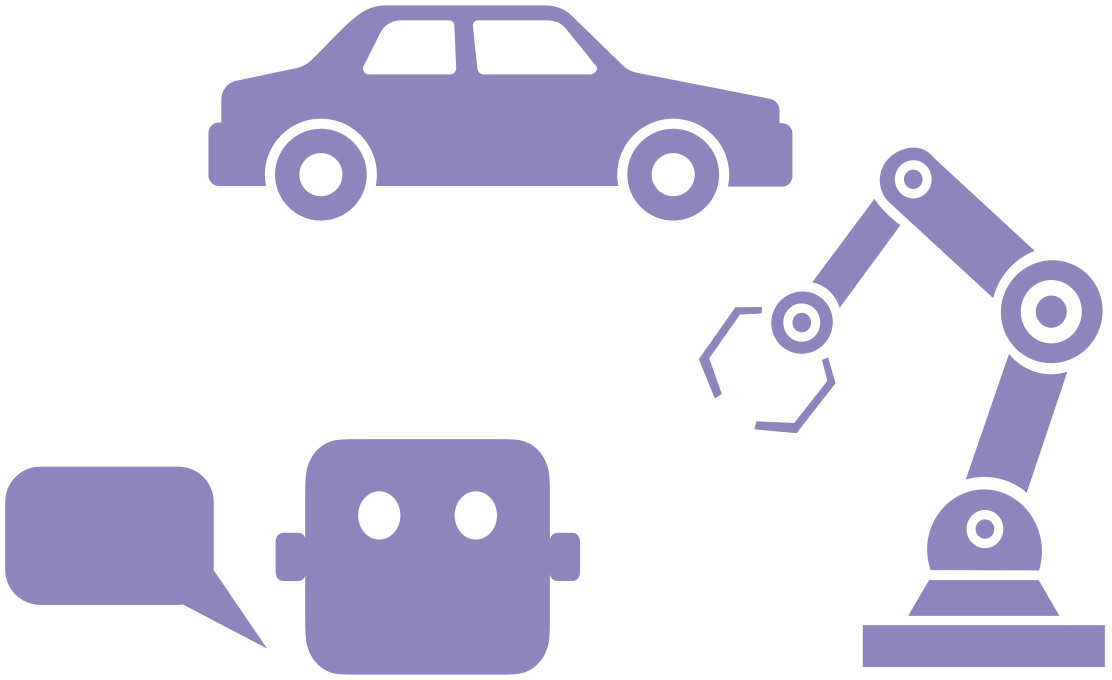} \\
\bottomrule
\end{tabular}

\end{sidewaystableplaceholder}

\subsection{Training: Learning a Generative Model}

In the last subsection, we discussed how in order to learn an unknown data distribution from samples, we need to implement an architecture of hierarchical dynamical systems. Importantly, the hierarchical architecture does not define a single distribution but a whole \emph{family} $\{p_\theta\}_\theta$, one for every setting of the parameters $\theta$. \emph{Training} is the process of selecting the set of parameters $\theta$ that makes $p_\theta$ as close as possible to the data distribution $p_{\mathrm{data}}$. In what follows, we provide an overview of training in generative models given their hierarchical architecture. Using the control-theoretical framework that we have established, we explain where in the architecture the learnable parameters live, provide objective functions that measure the closeness between the real and learned distribution, and review existing algorithms on how such objective is minimized. 

\paragraph{\textbf{Where the parameters live}} The learnable parameters are distributed across the maps of levels (i)--(iv). Level (v) is the exception: the interaction rule $\twom{\psi}$ is the external world, and is not ours to learn. Concretely, $\theta$ collects:
\begin{itemize}
  \item at the \one{token mixer}: the value projection $\onem{W_V}$ together
  with the mixing operator $\onem{M}$ in equation \eqref{eq:token_mixer} (parametrized by the query/key projections in attention, and the state matrices in SSMs);
  \item at the \five{mixer scaffold}: the $\gamma,\beta$ parameters of $\fivem{\text{Norm}}$, and the weights $W_1, W_2$ in the channel function $\fivem{g}$;
  \item at the \three{deep architecture}, the embedding and unembedding matrices, which map sequence tokens to vectors $s_k$ and back,  introduce training parameters as well. It is important to note that the decoding rule
$\threem{\phi}$ is a fixed sampling procedure.
\end{itemize}

Since each of the \four{layers} is made up of a \one{token mixer} scaffolded by its corresponding \five{mixer scaffold}, the depth axis introduces no parameters of its own. Often, the parameters of the \one{token mixer} and its corresponding \five{mixer scaffold} are referred to as \emph{layer weights}, $\theta=\{\theta^{(\ell)}\}_{\ell=1}^{L}$. When weights are tied, weights are shared across layers. 

In summary, a generative model captured in $\threem{\Phi}$ inherits most of its parameters from the layer stack $\fourm{F}$, along with the embedding and unembedding matrix. In modern models the resulting $\theta$ has between $10^9$ and $10^{12}$ entries.

\paragraph{\textbf{Solving an optimization problem}}

\emph{Training} is the process by which the parameters $\theta$ are fixed by solving the optimization problem where $p_\theta$ best matches the data distribution
$p_{\mathrm{data}}$, i.e.,
\begin{equation}\label{eq:training}
  \theta^\star \;=\; \arg\min_{\theta\in\mathbb R^P}\ \mathcal{L}(\theta),
\end{equation}
where the loss $\mathcal{L}$ measures the discrepancy between $p_\theta$ and
$p_{\mathrm{data}}$, and $P\sim 10^{12}$. The choice of loss function $\mathcal{L}$ depends on the choice of \three{deep architecture}, as described in sidebar \emph{Loss Function Types}.

\begin{sidebarcol}{Loss Function Types}
\begin{itemize}
  \item \textbf{Autoregressive~\cite{radford2018gpt}}: the negative
  log-likelihood, which factorizes along the token axis,
  \[
    \mathcal{L}(\theta) = -\,\mathbb{E}_{\mathbf{x}\sim p_{\mathrm{data}}}
    \Big[\textstyle\sum_{t=1}^{T}\log p_\theta\big(x_t\mid x_{<t}\big)\Big].
  \]
  \item \textbf{Diffusion / flow matching~\cite{ho2020ddpm,lipman2022flow}}: a
  denoising regression matching the backbone output to a target (a score or a
  velocity) at a randomly sampled step $k$,
  \[
    \mathcal{L}(\theta) = \mathbb{E}_{\mathbf{x},k}
    \big\|\,\fourm{F_\theta}(s_k,k) - \mathrm{target}(\mathbf{x},k)\,\big\|^2 .
  \]
  \item \textbf{Variational autoencoders~\cite{kingma2014vae}}: the (negative)
  evidence lower bound, a reconstruction term regularized by a KL term,
  \[
    \mathcal{L}(\theta) = -\,\mathbb{E}\big[\log p_\theta(\mathbf{x}\mid z)\big]
    + \mathrm{KL}\big(q(z\mid\mathbf{x})\,\|\,p(z)\big),
  \]
  where $\mathrm{KL}(\cdot\,\|\,\cdot)$ is the Kullback--Leibler divergence, a standard measure of how much one distribution differs from another, and $z$ is a latent variable with prior $p(z)$ and approximate posterior $q(z\mid\mathbf{x})$.
  \item \textbf{Generative adversarial networks~\cite{goodfellow2014gan}}: an
  adversarial objective, a min--max game against a discriminator $D$ rather than
  a fixed target,
\[
  \mathcal{L}(\theta) = \max_{D}\;
  \mathbb{E}_{\mathbf{x}\sim p_{\mathrm{data}}}\big[\log D(\mathbf{x})\big]
  + \mathbb{E}_{\tilde{\mathbf{x}}\sim p_\theta}\big[\log\big(1 - D(\tilde{\mathbf{x}})\big)\big].
\]
   This is the only one of the four with no likelihood interpretation at all. 
  \end{itemize}

\end{sidebarcol}

Despite this variety of objectives, they are all solved by the \emph{same}
algorithm: gradient descent. Starting from a random $\theta$, one
repeatedly draws a mini-batch of data, estimates the gradient
$\nabla_\theta\mathcal{L}$ on it by \emph{backpropagation} through the composed
maps of levels (i)--(iv), and takes a step
\begin{equation}\label{eq:sgd}
  \theta \;\leftarrow\; \theta - \eta\,\nabla_\theta\mathcal{L}(\theta),
\end{equation}
with learning rate $\eta$. In practice the bare update~\eqref{eq:sgd} is refined by adding \emph{state} to the optimizer: momentum accumulates a running mean of past gradients (a velocity), and the de facto standard, \emph{Adam}~\cite{kingma2015adam} (and its variant AdamW), additionally rescales each coordinate by a running estimate of the gradient's magnitude, giving a per-parameter adaptive step size. That this procedure reliably reaches a useful $\theta^\star$, over billions of steps and parameters, is what makes the entire generative model possible. 
Equation~\eqref{eq:sgd} is itself a discrete-time dynamical system on $\theta$: the slowest loop in the whole hierarchy, running offline at design time rather than at deployment. Even these dynamics occur at various temporal scales, usually referred to as \emph{training stages} and described below.

\paragraph{\textbf{Training Stages}}

In practice $\theta^\star$ is not reached in a single optimization but through a sequence of \emph{stages}, each solving problem~\eqref{eq:training} with a different loss and a different dataset. The stages move from generic to specific: from acquiring broad capabilities to shaping a particular behavior.

\begin{itemize}
    \item {\textbf{Pretraining.}} The first and by far largest stage solves \eqref{eq:training} with a self-supervised loss on a massive, generic corpus. During pretraining the bulk of $\theta$, and the model's general capabilities, are acquired. The result is a \emph{base model}: fluent and broadly knowledgeable, but not yet adapted to any particular task.
    \item \textbf{Supervised finetuning (SFT).} The base model is then adapted on a much smaller, curated set of input--output demonstrations, continuing the same likelihood loss but now on examples of the desired behavior. SFT does not teach new raw capability so much as elicit and format it: it shapes \emph{how} the model responds. Examples include following instructions and adopting a format or persona.
    \item \textbf{Reinforcement learning (RL).} The final stage optimizes $\theta$ against a scalar \emph{reward} rather than fixed targets, maximizing expected reward over the model's own generations, $\max_\theta\ \mathbb{E}_{\mathbf{x}\sim p_\theta}\big[\,R(\mathbf{x})\,\big]$, where the reward $R$ encodes human preferences (RLHF~\cite{instructgpt2022}) or an automatically verifiable signal (e.g.\ whether a proof checks or a test passes). Unlike the previous stages, the model here is optimized over \emph{its own outputs}: it is explicitly treated as a \emph{policy}, and improved through the interaction loop of \two{AI-world interactions}. 
\end{itemize}

\subsection{Inference: Predicting with Generative Models}

Once the model is trained, the parameters $\theta$ are frozen and the model is ready to generate content. \emph{Inference} refers to the generation of outputs through the architecture of hierarchical dynamical systems. As such,  producing an output is not a single parallel computation but a set of \emph{nested loops}: a single step along any axis requires running every axis below it to completion. In what follows, we illustrate this process starting with a prompt.

\paragraph{\textbf{0. From prompt to tokens.}} 

To begin an inference step, the generative model begins from a prompt (text, image, etc.) supplied by the user, which constitutes the initial context $c_0$ and which is embedded into $\mathbb{R}^{T\times d}$ to seed the first  generation state $s_0\in\mathbb{R}^{T\times d}$; and to run it through the backbone, that state is taken as the initial activation $h_0 = s_0$. This gives rise to a sequence of embedded tokens
\[
  h_0 \;=\; [\,x_1,\dots,x_T\,]^\intercal\;\in\;\mathbb{R}^{T\times d}.
\]
The computation now starts from these tokens.

\paragraph{\textbf{1. Token axis: inside one block.}} The token sequence is processed by a single block $\fivem{\mathcal{B}}[\onem{f}]$, whose operations apply in a fixed order. First the scaffold \five{normalizes} the tokens, the \one{token mixer} mixes information \emph{across} them, and a \five{skip connection} adds the result back,
\[
  z \;=\; h_0 \,+\, \onem{f}\big(\fivem{\mathrm{Norm}}(h_0)\big);
\]
then the same pattern repeats with the \five{channel function}, which recombines features \emph{within} each token,
\[
  h_1 \;=\; z \,+\, \fivem{g}\big(\fivem{\mathrm{Norm}}(z)\big).
\]
Of these operations only the \one{token mixer} moves information between positions; the rest act on each token independently. The block takes one activation $h_0$ and outputs the next, $h_1$.

\paragraph{\textbf{2. Depth axis: advancing} $h$ \textbf{through the layers.}} Stacking $L$ such blocks (\four{layers}) propagates the activation up the depth axis,
\[
  h_0 \;\to\; h_1 \;\to\;\cdots\;\to\; h_L \;=\; \fourm{F_\theta}(h_0).
\]
At every \four{layer} the activation is nothing but a sequence of token vectors, and the block above acts on it; so advancing the depth axis is precisely the token-level processing of the previous paragraph, \emph{repeated $L$ times}. One full sweep of this axis---a single \emph{forward pass}---creates a processed representation $h_L$ from the prompt's tokens.

\paragraph{\textbf{3. Generation axis: advancing} $s$ \textbf{through a forward pass.}}
A single forward pass does not, in general, finish the sample. The \three{decoding rule} outputs the next generation state from the final activation,
\[
  s_{1} \;=\; \threem{\phi}\big(s_0,\,\fourm{F_\theta}(s_0)\big),
\]
and this new state is then fed back as a fresh input activation, $h_0 \leftarrow s_1$, and sent down to (1) its tokens and back up through (2) the $L$ blocks once more. Advancing the \three{generation axis} a single step therefore requires running the entire depth axis to completion: $s$ is processed \emph{through} a forward pass $h$ each time. Iterating from $s_0$ to $s_K$ completes the sample, $\mathbf{x}=s_K\sim p_\theta$.

\paragraph{\textbf{4. Interaction axis: advancing the context via generations.}}
Once deployed, the finished sample is the model's \emph{action}, $a_0 = \threem{\Phi}(c_0) = s_K$. The \two{interaction rule} submits this action to the world, and the world's response is appended to the model's context yielding $c_1=\twom{\psi}(c_0,a_0)$; the next action then requires generating a whole new sample from that updated context. So advancing the \two{interaction axis} a single step requires running an entire generation to completion: the context is processed \emph{through} a generation $s$ each time, and grows as the interaction advances over time. 

\begin{sidebarcol} {Code Example: How to build a Transformer}

An interactive Colab notebook that builds each level of the hierarchy above as a swappable component, then trains and runs inference with the result:\\[4pt]
{\color{blue}\url{https://colab.research.google.com/drive/1uBdhrAl4FWgn3o98DFMWOd5Luie6OL-w?usp=sharing}}

\end{sidebarcol}

\section{Dynamics and Control at Training Time}

In this section, we present the dynamics arising during the training process. We note that, during training, the dynamics are those of the weight (parameter) updates:
\begin{equation}\label{eq:theta-update}
  \theta_{t+1} = \theta_t - \eta\,\nabla_\theta\mathcal{L}(\theta_t).
\end{equation}

Backpropagation routes the gradient through the token mixer~\eqref{eq:token_mixer}, the mixer scaffold~\eqref{eq:scaffold-types}, and the layer stack~\eqref{eq:layers} in turn, and each of these, evaluated at the current, frozen $\theta$, contributes a Jacobian that multiplies into $\nabla_\theta\mathcal{L}$. So the dynamical systems already defined for each axis do not reappear here as separate training-time trajectories; they reappear as \emph{gains} on the one true trajectory above, and the question at each axis is how well- or ill-conditioned that gain is. The five subsections below identify, for each axis, how its state maps into the corresponding Jacobian.

We also identify which axis matters at each training regime, since not every axis is exercised at every regime. The token and depth axes are traversed on every forward pass, in every stage of training, since the model can never be evaluated without mixing tokens and stacking layers. Generation and interaction have different techniques at pretraining, supervised finetuning, and RL stages.

\subsection{\one{In the Token Mixer}}

The token mixer is where $\theta$ is instantiated directly: in~\eqref{eq:token_mixer}, both the mixing operator $\onem{M(}\mathbf{x}\onem{)}$ and the value projection $\onem{W_V}$ are literal functions of $\theta$. Hence, the token axis is where a layer's local sensitivity to its own weights is computed by differentiating~\eqref{eq:token_mixer} itself,
\begin{equation}\label{eq:token-mixer-grad}
  \frac{\partial y_t}{\partial\theta} \;=\; \sum_{\tau=1}^{T}\left(\onem{W_V^\top}x_\tau\frac{\partial\,\onem{[M(}\mathbf{x}\onem{)]}_{t\tau}}{\partial\theta} \;+\; \onem{[M(}\mathbf{x}\onem{)]}_{t\tau}\,\frac{\partial\big(\onem{W_V^\top}x_\tau\big)}{\partial\theta}\right).
\end{equation}

Summed over token positions and weighted by the downstream loss gradient, this is exactly how the token axis enters $\nabla_\theta\mathcal{L}$,
\begin{equation}\label{eq:token-mixer-nabla}
  \nabla_\theta\mathcal{L} \;=\; \sum_{t=1}^{T} \frac{\partial\mathcal{L}}{\partial y_t}\,\frac{\partial y_t}{\partial\theta},
\end{equation}
with $\partial y_t/\partial\theta$ given by~\eqref{eq:token-mixer-grad}, and $\partial\mathcal{L}/\partial y_t$ carrying everything downstream of $y_t$ (the rest of the scaffold, the remaining layers, and the loss itself). The two terms inside~\eqref{eq:token-mixer-grad} then play different roles in this sum. The second, through $\onem{W_V}$, is a fixed linear map applied identically to every token pair: exactly as trainable at token distance $t-\tau$ as at distance zero. The first, through $\partial[M(\mathbf{x})]_{t\tau}/\partial\theta$, is where trainability across token distance actually lives: how strongly a training signal at $t$ reaches back to shape the parameters responsible for token $\tau$'s contribution depends entirely on how the mixing operator, and its sensitivity to $\theta$, behaves as $t-\tau$ grows. This is a structural property of how $M$ is parametrized, and sidebar \emph{Gradient Gain along Token Axis} makes this concrete for two common choices.

\begin{sidebarcol}[AppleGreen6!18!white]{Gradient Gain along Token Axis~\cite{ghodsi2025heads}}

As highlighted in \cite{ghodsi2025heads}, different structural choices for the token mixer can impact the dynamics during training. 
\begin{itemize}

	\item State-space model mixer induces a mixer operator of~\eqref{eq:token_mixer} \emph{implicitly} by unrolling a latent dynamical system, $J_{t\tau}:=\partial y_t/\partial x_\tau= C_t\Big(\prod_{j=\tau+1}^{t}\onem{A_j}\Big)\onem{B_\tau}$ is the resulting input--output Jacobian, which costs a state-space mixer its long-range gradient. Hence, since the transition is uniformly contractive ($\|\onem{A_j}\|\le\rho<1$), $\|J_{t\tau}\| \;\le\; \|C_t\|\,\|B_\tau\|\,\rho^{\,t-\tau}$, which vanishes exponentially in $t-\tau$.
	
	\item Attention factorizes the mixing operator $[M(\mathbf{x})]_{t\tau}$ of~\eqref{eq:token_mixer} \emph{explicitly}, as a scalar query--key score, $[M(\mathbf{x})]_{t\tau}=\sigma_\theta(x_t,x_\tau)$. Since it has no state to contract, it exhibits inputs for which $\|J_{t\tau}\|\ge\|W_V\|-\epsilon$ for any $\epsilon>0$, at \emph{any} distance $t-\tau$: a gradient path whose strength does not decay with distance.

\end{itemize}

\emph{Example of results (extracted from~\cite{ghodsi2025heads}):} Tracking the gradient norm $\|\nabla_{x_1}y_T\|$ against sequence length $T$ empirically confirms this asymmetry: the state-space mixer's gradient decays exponentially with $T$, while attention's decays far more slowly and remains substantially larger at long range.

\begin{center}
\includegraphics[width=0.7\linewidth]{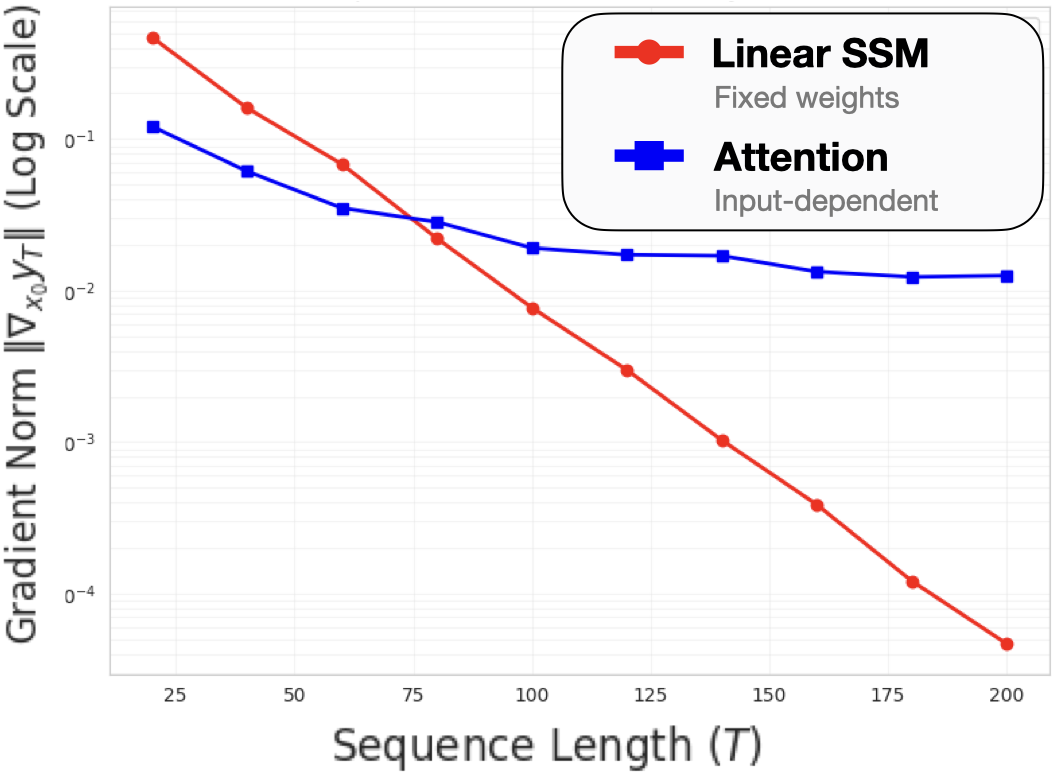}
\captionof{figure}{\footnotesize{Gradient norm $\|\nabla_{x_0}y_T\|$ (log scale) vs.\ sequence length $T$; reproduced from~\cite{ghodsi2025heads}. The linear SSM's gradient decays exponentially with $T$, while attention's decays far more slowly and stays substantially larger at long range.}}
\label{fig:gradient-highway-plot}
\end{center}

The control input native to this axis is $\onem{A_j}$ itself: making the transition data-dependent, as in Mamba's discretization $\onem{A_t}=\exp(\Delta_t A)$ \cite{gu2023mamba} gives direct authority over how contractive each step of it is, set per token rather than fixed once at initialization. This is why exploring the token axis in depth, as we do in the next section, is not purely an inference-time question: the same structural choices that shape $\onem{A_j}$ are exactly what would address this training-time control problem too.

\end{sidebarcol}

\subsection{\five{In the Mixer Scaffold}}

The mixer scaffold is what conditions the depth axis rather than adding dynamics of its own: in~\eqref{eq:scaffold-types}, the skip connection and the normalization are literal structural choices wrapped around the mixer $\onem{f}$ itself, and stacking $L$ such blocks,~\eqref{eq:layers}, is how the mixer becomes the trainable backbone $\fourm{F}$. Hence, the depth axis is where an early layer's local sensitivity to its own weights is computed by differentiating the backbone itself,
\begin{equation}\label{eq:depth-jacobian}
  \nabla_{\theta^{(\ell)}}\mathcal{L} = \frac{\partial\mathcal{L}}{\partial h_L}\,J_\ell\,\frac{\partial h_\ell}{\partial\theta^{(\ell)}}, \qquad J_\ell := \frac{\partial h_L}{\partial h_\ell} = \prod_{k=\ell+1}^{L}\frac{\partial h_k}{\partial h_{k-1}},
\end{equation}
so $J_\ell$ sets the size of every training-time step~\eqref{eq:sgd} takes for early-layer parameters. Each factor $\partial h_k/\partial h_{k-1}$ inside that product is the Jacobian of a single scaffold block's map $h_{k-1}\mapsto h_k$, so writing out that one map is exactly what $J_\ell$ is built from,
\begin{equation}\label{eq:ff-fb-depth}
  \underbrace{\tilde{h}_\ell = h_{\ell-1} + m_\ell}_{\text{feedforward}}, \qquad
  \underbrace{h_\ell = D(\tilde{h}_\ell)\,\tilde{h}_\ell}_{\text{feedback}},
\end{equation}
with $m_\ell$ the mixer's contribution at layer $\ell$; $\tilde{h}_\ell$ the intermediate activation produced by the feedforward step, and $D(\tilde{h}_\ell) = \mathrm{diag}(1/\|\tilde{h}_{\ell,i,:}\|_2)_{i=1}^{T}$ a correction that depends on $\tilde{h}_\ell$ itself. We identify two terms that ``control" the evolution: a \emph{feedforward} path, carrying the previous activation forward unconditionally with nothing to correct, and a \emph{feedback} path, measuring the very quantity it acts on and correcting it toward a fixed reference. Equation~\eqref{eq:ff-fb-depth} is post-norm: in the pre-norm block of~\eqref{eq:scaffold-types} the normalization acts only on the branch input, never on the stream.

We note, per~\eqref{eq:scaffold-types}, that the scaffold wraps this same feedforward-feedback pair around \emph{two} maps in sequence within a single block, the token mixer and the channel MLP, so the one factor $\partial h_k/\partial h_{k-1}$ above is itself the product of two such Jacobians, one per sub-block, chained together exactly as the $L$ layer-Jacobians are chained into $J_\ell$ in~\eqref{eq:depth-jacobian}. Since the choice of $\fivem{g}$ is architecture-dependent, in what follows we analyze the skip connection and the normalization. Both terms sit inside the same map $h_{\ell-1}\mapsto h_\ell$ that $J_\ell$ multiplies together in~\eqref{eq:depth-jacobian}: A single fixed recursion carries both corrections, so the skip connection and the normalization act along one and the same axis, depth, rather than each carving out an axis of its own. The two subsections below take each term of~\eqref{eq:ff-fb-depth} in turn.

\paragraph{\five{The role of skip connections}}
The skip connection in equation \eqref{eq:scaffold-types} is precisely the feedforward term in~\eqref{eq:ff-fb-depth}: $h_{\ell-1}$ passed forward with no correction. When the feedback term fails to correct what
this feedforward term lets through, every token's embedding is pushed toward the same shared vector as the number of layers increases -- known as \emph{rank collapse}~\cite{lambdaskip2025} -- and $J_\ell$ collapses (vanishes) with it. Collapse is measured by how far each row of $h_\ell$ sits from the sequence's mean,
\begin{equation}\label{eq:rank-collapse-measure}
  \fivem{\mu}(h_\ell) = \big\|h_\ell - \mathbf{1}\gamma_{h_\ell}\big\|_F, \qquad \gamma_{h_\ell} = \tfrac{1}{T}\mathbf{1}^{\top}h_\ell.
\end{equation}
Sidebar \emph{Skip Strength as a Control Gain} turns this collapse rate into an explicit, tunable condition on the skip connection itself.
\begin{sidebarcol}[AppleRed5!18!white]{Skip Strength as a Control Gain~\cite{lambdaskip2025}}

As explored in \cite{lambdaskip2025}, the skip connection's strength can be seen as an explicit, tunable gain $\fivem{\lambda}$, i.e., $\tilde{h}_\ell = \fivem{\lambda}\, h_{\ell-1} + m_\ell$, recovering the standard skip connection at $\fivem{\lambda}=1$. Being purely feedforward, $\fivem{\lambda}$ enters $J_\ell$ in~\eqref{eq:depth-jacobian} additively and multiplicatively: it has no reference to correct toward, so it can only ever scale the training-time step~\eqref{eq:sgd} takes for early layers. Without a skip connection at all ($\fivem{\lambda}=0$), $\fivem{\mu}(h_L)$ collapses exponentially in depth $L$, and gradients reaching the earliest layers vanish at the same rate.
Writing the collapse rate as $a>0$ -- the rate at which the lower bound on $\fivem{\mu}(h_L)^2$ decays with depth -- $\fivem{\lambda}$ is what sets that rate: whenever
\begin{equation}\label{eq:lambda-condition}
  \fivem{\lambda}^2 \;>\; a\big(SC_M+|\fivem{\lambda}|\big)^2,
\end{equation}
with $C_M$, $S$ bounding the mixer and value map, $\fivem{\mu}(h_L)^2 \ge a^L\fivem{\mu}(h_0)^2$ for every depth $L$~\cite{lambdaskip2025}. The condition is feasible only for $a < 1$, but $a$ can be driven arbitrarily close to one by increasing $\vert \fivem \lambda \vert$, so the collapse is made arbitrarily slow (and negligible at any finite depth) for any architecture expressible in this form. This is exactly the reading~\eqref{eq:depth-jacobian} already gives $\fivem{\lambda}$: a multiplier on the training-time update: a gain, chosen once and fixed for the whole architecture.

\emph{Example of results (extracted from~\cite{lambdaskip2025}):} sweeping $\fivem{\lambda}$ post-hoc through a frozen, pretrained Transformer (Albert) and SSM (Mamba) checkpoint and measuring the normalized rank-collapse metric at the last layer confirms this: for both architectures, $\fivem{\mu}$ collapses sharply in a narrow band around $\fivem{\lambda}=0$, and recovers as $|\fivem{\lambda}|$ grows in either direction, exactly as~\eqref{eq:lambda-condition} predicts.

\begin{center}
\includegraphics[width=0.85\linewidth]{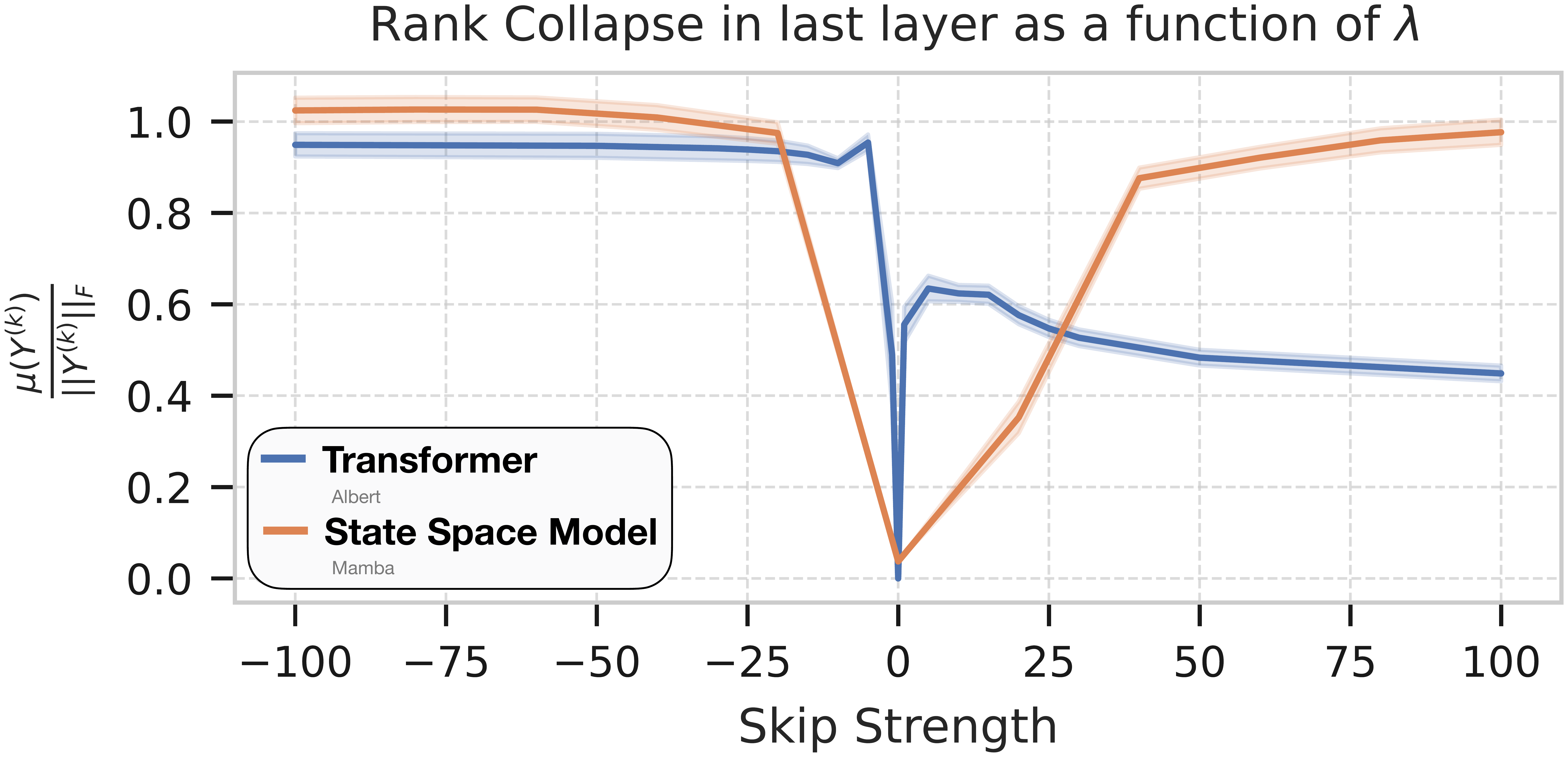}
\captionof{figure}{\footnotesize{Normalized rank-collapse measure $\mu(h_L)/\|h_L\|_F$ at the last layer vs.\ skip strength $\lambda$, for a Transformer (Albert) and a State Space Model (Mamba); reproduced from~\cite{lambdaskip2025}. Both architectures collapse sharply near $\lambda=0$ and recover as $|\lambda|$ grows in either direction.}}
\label{fig:lambda-skip-plot}
\end{center}

Further exploring the depth axis, as we do in the next section, leads to a generalization of this idea: from a fixed scalar gain to a genuine feedback law, $\tilde{h}_\ell = h_\ell + u_\ell(h_\ell)$, where the correction is computed from the current activation rather than baked in at training time~\eqref{eq:layers-control}: the same structural idea that stabilizes training here is what makes post-hoc, training-free control possible there.

\end{sidebarcol}

\paragraph{\five{The role of normalization}}
Normalization is precisely the feedback term in~\eqref{eq:ff-fb-depth}, $D(\tilde{h}_l)$, and unlike the skip connection it plays this role twice. Across depth, on its own, without a skip connection, it never \emph{prevents} rank collapse for any of the architectures considered in~\cite{lambdaskip2025}, but it does slow its \emph{rate}~\cite{lambdaskip2025} 
. But normalization also has a second role. Left uncorrected, gradient descent inflates the scale $\fivem{r}(h)=\sqrt{\|h\|_2^2/d+\epsilon}$ of the final representation on its own since on any already-correct token, a longer $h$ sharpens the softmax and lowers the loss for free. This unforced, one-directional drift is known as \emph{logit chasing}. Tracking $\fivem{r}(h)$ across training steps $t$ in the gradient-flow limit gives
\begin{equation}\label{eq:scale-dynamics}
  \fivem{\dot{r}} \;=\; \underbrace{-\frac{1}{d\,\fivem{r}}\big\langle\nabla_h\ell,\,h\big\rangle}_{\text{uncontrolled forcing}} \;+\; \underbrace{u(\fivem{r})}_{\text{feedback, if present}},
\end{equation}
with $\ell_{tok}$ the per-token loss inside $\mathcal{L}$. When the margin is positive, i.e., the correct logit exceeds the probability-weighted average logit, scaling up $h$ lowers the loss, so $\langle\nabla_h\ell_{tok},h\rangle<0$. Hence, without a feedback term $u(\fivem{r})$, the scale grows unchecked, $\fivem{\dot{r}}>0$.

\subsection{\four{In the Layers}}

The layer stack does not introduce a new mechanism beyond the one already identified in the mixer scaffold: the depth-axis Jacobian $J_l$ in~\eqref{eq:depth-jacobian} is already the product over \emph{all} $L$ layers, so once the skip connection and the normalization are fixed, what is left to ask is the empirical question of how that product actually distributes across depth once training converges. Pre-norm stacks answer this consistently: the gradient reaching layer $l$ is the sum of the identity (feedforward) path in~\eqref{eq:ff-fb-depth} and every downstream functional (feedback) path, and this sum shrinks with depth, so early layers receive systematically richer gradients than late ones. This is a known structural asymmetry that follows directly from~\eqref{eq:ff-fb-depth} itself~\cite{cascadeformer2026}. The natural symptom of this asymmetry is what empirical work independently reports: large blocks of deep layers can be removed from a pretrained network with only minor loss in performance, and removability increases with depth up to the final few layers, which remain essential~\cite{gromov2024ineffectiveness}. Both findings corroborate, rather than extend, the picture already built above: a poorly-conditioned $J_l$ does not just slow training, it leaves late layers undertrained, and the resulting redundancy is diagnosed after the fact rather than corrected during training itself. Whether that gradient allocation can be corrected online remains an open question.

\subsection{\three{In the Deep Architecture}}

The deep architecture is where the generation axis is instantiated directly: in~\eqref{eq:autonomous-generation}, iterating the backbone $\fourm{F_\theta}$ through the decoding rule $\threem{\phi}$ for $K$ steps is what turns a single forward pass into the generation loop itself. Before turning to the generation axis itself, sidebar \emph{Normalization as Scheduled Control} closes the open question left by~\eqref{eq:scale-dynamics}: what supplies the missing feedback term $u(\fivem{r})$.

\begin{sidebarcol}[AppleRed5!18!white]{Normalization as Scheduled Control~\cite{tapernorm2026}}

Normalization is what can supply $u(\fivem{r})$: unlike a fixed rescaling, normalization recomputes at every forward pass its correction from the value of $h$. That same adaptivity is also what makes it expensive: standard normalization is a nonlinear, per-token computation, paid at every layer and every token during both training \emph{and} inference. That raises a question: could its adaptivity be phased out on a schedule instead, staying fully active while $\fivem{r}(h)$'s drift is largest and fading out only once the rest of training can compensate for a fixed correction in its place? \cite{tapernorm2026} answers this question by supplying the missing $u(\fivem{r})$ with a gate $\fivem{\kappa_t}\in[0,1]$, itself scheduled along the training axis: held at $\fivem{\kappa_t}=1$ through an initial warmup, recovering ordinary normalization, then cosine-decayed to $\fivem{\kappa_t}=0$ over the remainder of training,
\[
  \mathrm{TaperNorm}(\tilde h_\ell;\fivem{\kappa_t}) \;=\; \fivem{\kappa_t}\,D(\tilde h_\ell)\tilde h_\ell \;+\; (1-\fivem{\kappa_t})\,c\,\tilde h_\ell D_{\tilde\gamma},
\]
interpolating between the adaptive correction $D(\tilde h_\ell)\tilde h_\ell$ itself and a fixed, sample-independent affine map, with $c\in\mathbb{R}$ and $D_{\tilde\gamma}$ frozen once tapering starts.
By $\fivem{\kappa_t}=0$ the correction no longer depends on the current activation at all, so it can be folded into the adjacent linear projection. Empirically, at inference, the efficiency payoff is up to $1.18\times$ higher inference throughput.
It is important to note that while tapering internal normalization costs little -- \cite{tapernorm2026} reports roughly a $1$--$2\%$ relative increase in validation loss from removing it -- the anchoring provided by the normalization at the last layer is not something a schedule alone can retire: the final layer must either stay adaptive, or be explicitly substituted.

Further exploring strategies that actively control relevant quantities during training dynamics offers great promise to replace existing components by more principled, efficient and effective ones.

\end{sidebarcol}

\noindent
Hence, the generation axis is where the loss's dependence on $\theta$ is decided by how many of these $K$ steps training actually chains together,
\begin{equation}\label{eq:generation-jacobian}
  \begin{gathered}
  \nabla_\theta\mathcal{L}\big|_{k} = \frac{\partial\mathcal{L}}{\partial s_K}\,J_{\text{gen},k}\,\frac{\partial s_{k+1}}{\partial\theta}\bigg|_{s_k}, \\[3pt]
  J_{\text{gen},k} := \frac{\partial s_K}{\partial s_{k+1}} = \prod_{j=k+2}^{K}\frac{\partial s_j}{\partial s_{j-1}},
  \end{gathered}
\end{equation}
with $J_{\text{gen},k}$ the generation-axis analogue of the depth-axis Jacobian $J_\ell$ in~\eqref{eq:depth-jacobian}, chaining every remaining step of the loop from $k+1$ to $K$. Whether, and how, training ever accounts for this closed-loop Jacobian -- by avoiding it outright, or by finding a gradient that does not require it -- depends on what is actually fed to $\fourm{F_\theta}$ at each step, which is the subject of what follows.

\paragraph{\three{In the pretraining and SFT training stages}}
Along the generation axis, both pretraining and supervised finetuning induce an \emph{open-loop} training scheme. The most popular approach is \emph{teacher forcing}: at every step $k$, the ground-truth prefix $s_k^{\mathrm{gt}}$ is substituted for whatever the model's own rollout $\hat s_k := \threem{\phi}\big(s_{k-1}, \fourm{F_\theta}(s_{k-1})\big)$ would have produced, so the loss $\mathcal{L} = \sum_k \ell_{tok}\big(\fourm{F_\theta}(s_k^{\mathrm{gt}}), x_{k+1}^{\mathrm{gt}}\big)$ is always computed against data as the model's own generation never enters the loss. The idea is not specific to autoregressive, token-by-token decoding: \emph{diffusion forcing}~\cite{chen2024diffusionforcing} generalizes it to a continuous per-token corruption level, training each token to be recovered from an independently noised version of itself rather than from a fully clean or fully autonomous prediction. In control terms, this substitution is a feedback correction $u_k := s_k^{\mathrm{gt}} - \hat s_k$ layered on top of the autonomous update, but one that directly overrides the state completely, resetting $s_k$ to the reference regardless of what the autonomous dynamics produced. This is a \emph{dead-beat} correction, in the control-theoretic sense of a feedback law that drives the tracking error to zero in a single step, and it is precisely this property that severs $\partial s_j/\partial s_{j-1}$ in~\eqref{eq:generation-jacobian}: a correction with no dependence on the state it replaces has, by construction, zero sensitivity to that state.

Teacher forcing is widely used for its stability against exploding and vanishing gradients, and for its parallelizability -- every $s_k^{\mathrm{gt}}$ is available at once, so there is no $K$-step rollout to run. This comes at the cost of robustness: training never sees, let alone corrects for, how an error made at step $k$ would have propagated into the model's own rollout at $k+1$ -- known as \emph{exposure bias}. Once the model's own generation departs even slightly from the ground-truth manifold it was trained on, it is confronted with inputs it was never trained to correct, and small early errors can snowball over the remaining steps -- a failure mode characterized explicitly for diffusion decoders as well~\cite{ning2024elucidating,ren2024mdss}. Two other strategies address this gap directly. \emph{Free running} drops the correction altogether, $s_k := \hat s_k$ for every $k$ (i.e.\ $u_k \equiv 0$), training end to end on the model's own rollout. The drawback is that gradients are prohibitively noisy early in training, when $\fourm{F_\theta}$'s own predictions are still close to random. \emph{Scheduled sampling}~\cite{scheduledsampling2015} splits the difference, annealing a stochastic curriculum from teacher forcing toward free running over the course of training, so the model is only asked to correct its own rollout once it has something worth correcting. Sidebar \emph{Teacher Forcing and System Identification} recasts this entire exposure-bias story in the language of classical system identification.

\paragraph{\three{In the RL training stage}}
RL post-training does not supply a per-step reference the way teacher forcing does. In fact, it structurally could not because the reward $R(s_K)$ that scores a generation is a property of the completed rollout, not of any single step, so no target $s_k^{\mathrm{gt}}$ is available until the sequence is finished. RL is therefore necessarily a rollout-level regime, correcting a complete, sampled trajectory rather than one step at a time. Nor does it form $J_{gen,k}$ in~\eqref{eq:generation-jacobian} literally: the decoding rule $\threem{\phi}$ samples a token at every step, and no pathwise gradient runs through that sampling operation. Policy-gradient methods~\cite{ahmadian2024basics} sidestep this rather than solve it, trading the (unavailable) pathwise gradient through the dynamics for a gradient through the trajectory's own likelihood,
\begin{equation}\label{eq:rl-generation-grad}
  \begin{gathered}
  \nabla_\theta \mathcal J(\theta) = \mathbb{E}\Big[\nabla_\theta \log \pi_\theta(s_K)\,\big(R(s_K) - \threem{\beta}\text{log}\frac{\pi_\theta(s_K)}{\pi_{\mathrm{ref}}(s_K)}\big)\big)\Big], \\[3pt]
  \log\pi_\theta(s_K) = \sum_{k=0}^{K-1}\log\pi_\theta\big(x_{k+1}\mid s_k\big),
  \end{gathered}
\end{equation}
where $\pi_\theta$ is the distribution over trajectories induced by~\eqref{eq:autonomous-generation}, and $\pi_{\mathrm{ref}}$ is a reference policy (typically the teacher-forced model of the previous paragraph). Although the sum defining $\log\pi_\theta(s_K)$ is computed one token at a time, every term in it is weighted by the same scalar $R(s_K)$: credit is assigned to the whole trajectory at once. This is the opposite of teacher forcing's dead-beat correction: teacher forcing corrects every step immediately, independently, and with a per-token reference, whereas RL corrects the trajectory as a whole, only once it is complete, and with no reference beyond the reward itself.

\begin{sidebarcol}[AppleOrange5!18!white]{Teacher Forcing and System Identification}

System identification distinguishes \emph{open-loop} identification, where a model is fit on data collected without feedback, from \emph{closed-loop} identification, where the fitting data reflects the system operating under the very feedback law the model will be deployed with; a foundational finding is that a model fit open-loop can be arbitrarily poor precisely in the operating regime closed-loop control needs it to be accurate in, even while its open-loop prediction error looks small~\cite{vandenhof1995,hjalmarsson1996}. Exposure bias is this same result, restated for generative models: teacher forcing is open-loop identification of $\fourm{F_\theta}$, and autoregressive inference is the closed-loop operation the identified model is then asked to support. Seen this way, free running and scheduled sampling are informal, heuristic attempts at closed-loop identification without the formal guarantees, persistency-of-excitation conditions, and bias-variance analyses that closed-loop identification theory was built to provide for exactly this mismatch. Bringing that apparatus to bear on a $K$-step, high-dimensional, nonlinear generation loop remains, to our knowledge, entirely open.

\end{sidebarcol}

\subsection{\two{In AI-World Interactions}}

AI-world interactions are where the interaction axis is instantiated directly: iterating the interaction rule $\twom{\psi}$ against the generation-axis policy $\threem{\Phi}$ in~\eqref{eq:interaction} across $N$ exchanges is what turns a single action into the interaction loop itself. As along the generation axis, whether training accounts for this closed loop depends entirely on what supplies $a_n$ at each step, which is the subject of what follows.

\paragraph{\two{In the pretraining and SFT training stages}}
The interaction axis has its own open-loop regime, and its own name for it: \emph{behavior cloning}, or imitation learning from expert demonstrations, is the interaction-axis regime of the SFT stage. At every step $n$, the demonstrated action $a_n^{\mathrm{gt}}$ is substituted for whatever the policy's own rollout $\hat a_n := \threem{\Phi}(c_n)$ would have produced, so training regresses directly onto expert data rather than onto the consequences of the policy's own choices. This is the interaction-axis analogue of teacher forcing, down to the override itself: $a_n^{\mathrm{gt}}$ replaces $\hat a_n$ regardless of what the policy proposed. The same trade-off follows. Behavior cloning is stable and simple to train, but because the world's response $c_{n+1} = \twom{\psi}(c_n, a_n)$ is never evaluated on the policy's own actions during training, the policy never learns to correct for small deviations once it starts acting on its own: the context distribution the demonstrations were drawn from and the one the policy actually visits at deployment drift apart, and the gap compounds with every step -- known as \emph{covariate shift}. To address this gap, dataset aggregation
(DAgger) \cite{ross2011dagger} queries the expert for the correct action \emph{on the states the learner's own policy actually visits}, aggregates this into the training data, and retrains,  rather than fixing the training set in advance. This is a heuristic, iterative attempt at collecting training data under the very feedback law the policy will be deployed with, in the closed-loop-identification sense.

\paragraph{\two{In the RL training stage}}
Unlike imitation learning, RL closes the loop in the interaction dynamics during training since the parameter update is computed from the model's own actions $a_n$ once they have been scored. Because the true objective (what a human ultimately wants) is not directly observable, training substitutes a learned proxy $\twom{R_\phi}$ for it, most often a reward model fit to a dataset of preferences,

{\footnotesize
\setlength{\abovedisplayskip}{\baselineskip}
\setlength{\belowdisplayskip}{\baselineskip}
\setlength{\abovedisplayshortskip}{0pt}
\setlength{\belowdisplayshortskip}{0pt}
\begin{equation}\label{eq:interaction-reward-grad}
  \nabla_\theta \mathcal{J}(\theta) = \mathbb{E}\Big[\nabla_\theta\log\pi_\theta(a_n\mid c_n)\,
    \Big( \twom{R_\phi}(c_n,a_n)
      - \threem{\beta}\,\log\tfrac{\pi_\theta(a_n\mid c_n)}{\pi_{\mathrm{ref}}(a_n\mid c_n)}\Big)\Big].
\end{equation}}

\noindent This gradient takes the same form as the generation-axis gradient of~\eqref{eq:rl-generation-grad}, replacing $s_K$ with the interaction loop's own action $a_n$. The gap between $\twom{R_\phi}$ and the true objective it stands in for is the signal driving $\theta$ along this axis. Sidebar \emph{Reward as an Imperfect Sensor of the World} examines what happens once that gap is exploited.

\begin{sidebarcol}[ApplePurple5!18!white]{Reward as an Imperfect Sensor of the World}

A learned reward $\twom{R_\phi}$ is only a proxy for the true objective, and it can fail in two ways: it can go stale or be exploited. \cite{nsdpo2024} models preference drift as a \emph{non-stationary} reward: rather than treating preferences as fixed, they let the reward function drift with time and show, in theory and in finetuning experiments, that simple exponential down-weighting of stale preference data has bounded regret, whereas standard DPO degrades. \cite{rewardhacking2026} documents the exploitation case: their setting is an RL coding task scored by program execution rather than by a learned $\twom{R_\phi}$, in which the model can rewrite the evaluator. Hacking is attempted from the outset and fails; the policy then falls back on solving; and it returns to hacking once the shortcut pays better than further improvement.

Both failure modes are analogous to the unforced drift already identified along the training axis for the final representation's scale in~\eqref{eq:scale-dynamics}: an uncorrected quantity (there $\fivem{r}(h)$, here the gap between what the reward measures and what we actually want) moves in whatever direction lowers the training loss fastest, with nothing in the objective correcting it back. One mitigation, used there and in~\eqref{eq:rl-generation-grad}, is an explicit anchor, here the KL penalty against a reference policy, trading off how far the policy may drift from a trusted baseline against how much reward it is allowed to chase; it limits drift but does not repair a stale or exploitable reward, which is why \cite{nsdpo2024} and \cite{rewardhacking2026} each add a remedy of their own.

\end{sidebarcol}

\section{Dynamics and Control at Inference Time}
In this section, we present the dynamics arising in the different axes during the generation process (inference time). We model as a dynamical system the token mixer (token axis), the layers (depth axis), the deep architecture (generation axis), and the AI-world interaction (interaction axis). We also explore different examples of what a control theoretic understanding of these systems can bring.

\subsection{\one{In the Token Mixer}}

Operating at the token axis, the token mixer in its convolutional form is captured in equation~\eqref{eq:token_mixer}. While this formulation might create the impression that the token mixer is a static forward pass, we now show how the token mixer introduces dynamics along the token axis. 

\paragraph{\one{The dynamics in the token axis}}
Before any tokens can be mixed, each one is projected into three separate signals
through learned matrices,
\begin{equation}\label{eq:qkv}
  q_t = \onem{W_Q^{\top}}x_t, \qquad
  k_t = \onem{W_K^{\top}}x_t, \qquad
  v_t = \onem{W_V^{\top}}x_t,
\end{equation}
the \emph{query}, the \emph{key}, and the \emph{value}, each playing a distinct role. The value $v_t\in\mathbb{R}^{d}$ is the information the token contributes to the others, i.e., the content that is actually passed along and combined. The key $k_t\in\mathbb{R}^{m}$ is a descriptor that advertises what the token offers, so that other tokens can judge how relevant it is. The query $q_t\in\mathbb{R}^{m}$ expresses what the token is currently looking for in the rest of the sequence. Mixing tokens amounts to letting each position gather the values of the others in proportion to how well its query matches their keys. Concretely, the output at position $t$ is a linear combination of the (past)
values,
\begin{equation}\label{eq:mixer_lincomb}
  y_t \;=\; \sum_{\tau\le t}\alpha_{t\tau}\,v_\tau,
\end{equation}
where the coefficient $\alpha_{t\tau}$ (how much token $t$ draws from token
$\tau$) is precisely the entry $\onem{[M(}\mathbf{x}\onem{)]_{t\tau}}$ of the mixing operator in equation~\eqref{eq:token_mixer}. The token dynamics arise from the fact that \emph{coefficients are not produced in one shot, but are the output of a dynamical system along the token axis}~\cite{sieber2025design}.
To see this, read each coefficient as the response of a small system that receives
the key $k_\tau$ as an \emph{impulse} at position $\tau$, evolves, and is read out
by the query $q_t$ at position $t$:
\begin{equation}
\label{eq:mixer_coeff}
\begin{aligned}
\xi_{t,\tau}
&= \onem{A_t}\,\xi_{t-1,\tau}
   + \onem{b_t}\,u^{(\tau)}_t, \quad
\alpha_{t\tau}
= \frac{1}{\nu_t}\onem{\rho}\!\left(q_t^{\top}\xi_{t,\tau}\right), \\
u^{(\tau)}_t
&=
\begin{cases}
k_\tau, & t=\tau,\\
0, & t\neq\tau.
\end{cases}
\end{aligned}
\end{equation}
where $\xi_{t,\tau}\in\mathbb{R}^{m}$ is the hidden state seeded by the impulse at $\tau$ (with $\xi_{\tau-1,\tau}=0$); $\onem{A_t}\in\mathbb{R}^{m\times m}$ is the (in general data-dependent) state transition; $\onem{b_t}\in\mathbb{R}$ scales the impulse; and $\onem{\rho}:\mathbb{R}\to\mathbb{R}$ is the \emph{readout map}. The value $v_\tau$ stays \emph{outside} this recurrence and is attached only at recombination~\eqref{eq:mixer_lincomb}, after a normalization $\nu_t$. Solving the recurrence from its zero initial state propagates the impulse from $\tau$ to $t$ through the product of transition matrices, and the convolutional form has the following entries:
\begin{equation}\label{eq:mixer_unroll}
  \onem{[M(}\mathbf{x}\onem{)]_{t\tau}} \;=\;
  \frac{1}{\nu_t}\,\onem{\rho}\!\Big(q_t^{\top}\Big(\textstyle\prod_{j=\tau+1}^{t}\onem{A_j}\Big)\onem{b_\tau}\,k_\tau\Big).
\end{equation}
Each entry is the gain from token $\tau$ to token $t$ accumulated by the dynamics between them: the static matrix of the architecture section is the \emph{unrolled trajectory} of system~\eqref{eq:mixer_coeff}. 

\begin{sidebarfig}[AppleGreen6!18!white]
    \includegraphics[width=\linewidth]{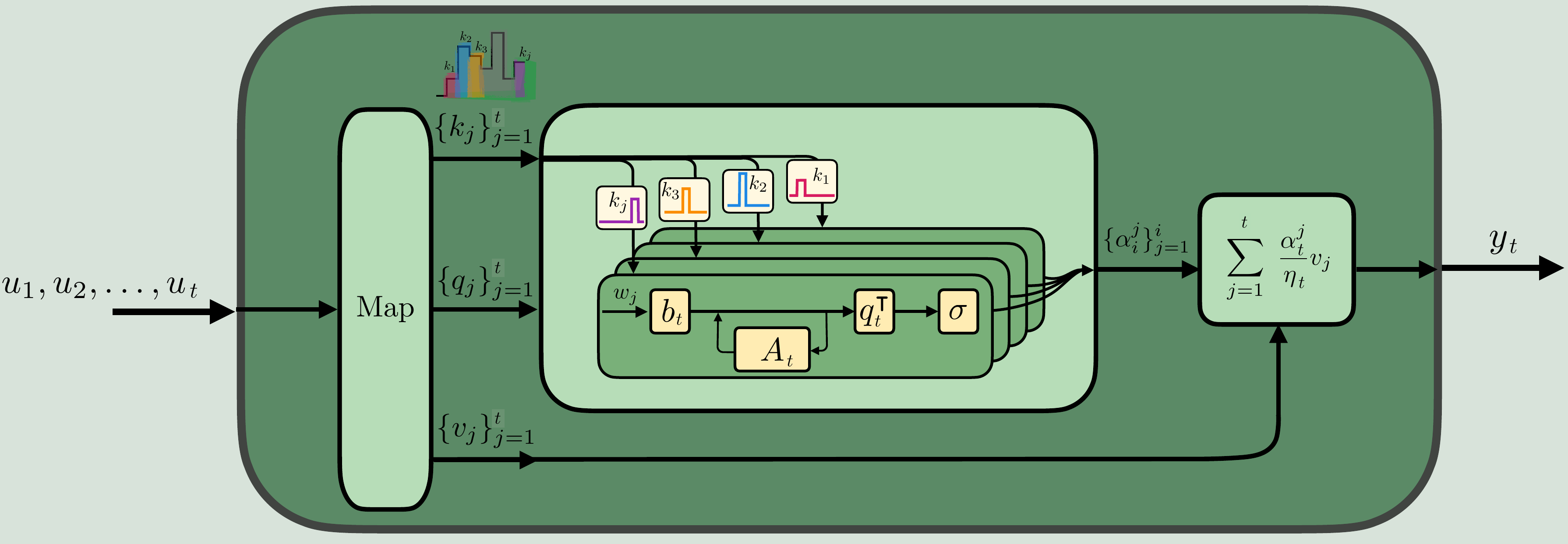}
    \caption{\footnotesize{ The sequence of token inputs are projected into three different signals: keys, queries, and values. The queries appear as a linear readout map of a dynamical system with matrices $A_t$, $b_t$, and a nonlinear readout $\rho$. We have $t$ copies of each of these dynamical systems per given $q_t$. Each copy gets excited with a key signal as an impulse excitation. Hence, the output of the system is the weighted sum of the values, where each coefficient is the impulse response of the corresponding dynamical system.}}
    \label{fig:my_label}
\end{sidebarfig}

\paragraph{\one{Control for the dynamics in the token axis}}

Equation~\eqref{eq:mixer_coeff} is not a single fixed map but a template with different elements subject to design choices: the state transition $\onem{A_t}$, which governs how the impulse from token $\tau$ propagates forward; the impulse scaling $\onem{b_t}$, which sets how strongly each key enters the state; the readout $\onem{\rho}$, which turns the query--state interaction into a coefficient; and the normalization $\onem{\nu_t}$, which rescales the coefficients before recombination. In this way, equation~\eqref{eq:mixer_coeff} can be seen as a closed-loop dynamical system. Hence, much of the power of the dynamical-systems view is that it collects what otherwise look like unrelated architectural tricks into these four interpretable components of one system ~\cite{sieber2024understanding,sieber2025design}. In the sidebar \emph{Design Principles for Token Mixer Dynamics}, we give an intuition for these design principles, which are formalized in~\cite{sieber2025design}. Sidebar \emph{Alternative View of the Token Axis Dynamics} gives an equivalent, value-augmented state representation of the same~\cite{sieber2024understanding}.

\begin{sidebarcol}{Alternative view of the token axis dynamics}

An alternative view of the dynamics exists~\cite{sieber2024understanding}, where values are directly included into the state. That is, the recurrence is no longer carried by the transformed key alone but by the key--value outer product, and one recovers the alternative representation in which the state has dimension $m\cdot d$ and the values live \emph{inside} the dynamics rather than outside. The readout map is the only obstacle between the two views: when it is linear, the transformation is straightforward. However, in the nonlinear case (such as in softmax attention~\cite{vaswani2017}), lifting the dimension of the state is often necessary, sometimes up to infinite-dimensional systems~\cite{sieber2024understanding}. The two descriptions are therefore the same system at two state sizes, trading \emph{readout nonlinearity} against \emph{state dimension}; the dial between them is the polynomial degree of $\onem{\rho}$, since a degree-$Q$ readout collapses into a linear-readout recurrence whose state expands as $m^{Q}$. Softmax attention, with $\onem{\rho}=\exp$, sits at the end of this dial: its lift is infinite-dimensional~\cite{sieber2024understanding}, so no finite recurrence exists,  coefficients are formed pairwise, and cost is $\mathcal{O}(T^2)$.

\end{sidebarcol}

\begin{sidebar}[AppleGreen6!18!white]{Design Principles for Token Mixer Dynamics~\cite{sieber2025design}}

\begin{itemize}
  \item \textbf{Readout $\onem{\rho}$:} A polynomial readout admits a finite recurrence with efficient implementation, whereas a non-polynomial one (as in softmax attention) does not. At the same time, the readout controls \emph{input selectivity}: the larger the region on which $\onem{\rho}$ is (near) zero, the more robustly irrelevant tokens can be suppressed. For instance, $\exp$ (near-zero values over the whole negative axis) selects sharply, while a linear readout (zero only at the origin) suppresses fragilely and needs to be compensated with specific choices of the transition matrix.
  \item \textbf{Transition $\onem{A_t}$:} Its structure sets which operations the dynamics perform on the keys: scalar transitions scale them, diagonal ones scale per coordinate, Householder ones rotate; and confining its eigenvalues to the unit disk guarantees stable, gradient-safe propagation along the token axis.
  \item \textbf{Impulse scaling $\onem{b_t}$:} Choosing $\onem{b_t}=\mathcal{O}(1/\sqrt{m})$ keeps the query--state product at unit variance regardless of state size; larger scalings let the variance grow with $m$ and destabilize training. This is the dynamical-systems generalization of the familiar inverse-square-root scaling of attention scores~\cite{vaswani2017}.
  \item \textbf{Normalization $\onem{\nu_t}$:}
  When the readout is unbounded or the transition is unstable, $\onem{\nu_t}$ must grow with the coefficients to keep them bounded; and the canonical choice $\onem{\nu_t}=\sum_{\tau\le t}\alpha_{t\tau}$ (softmax's denominator) forces the coefficients to sum to one, so that the output is a \emph{convex combination} of
  values, confined to their convex hull.
\end{itemize}

Standard choices for token mixer design are as follows in well-known architectures. For example, \textbf{Attention}~\cite{vaswani2017} has a nonlinear readout $\onem{\rho}=\exp$ with a near-integrator transition $\onem{A_t}\!\approx\!I$ (no forgetting); unbounded effective state, maximal expressivity, $\mathcal{O}(T^2)$ cost. It is important to remark that Attention is never implemented as a recurrence, its unrolled version leads to a direct implementation of the attention matrix. \textbf{Recurrent networks}~\cite{hochreiter1997lstm} implement a linear readout, a finite state, and a data-dependent diagonal forget gate $\onem{A_t}$ with eigenvalues inside the unit disk. \textbf{State-space models (S6)}~\cite{gu2023mamba} exhibit a linear readout, a finite state, and an input-dependent stable transition $\onem{A_t}=e^{-\Delta_t A}$; linear recurrence with a parallel scan.

\emph{Example of results (directly extracted from~\cite{sieber2025design}):} For a model to \emph{ignore} a token, its coefficient $\alpha=0$. This means that the impulse response has to be mapped to zero by $\onem{\rho}$. Readouts that have a large (near) zero level set are easier to map ``irrelevant'' query--key pairing to such set and get forgotten. 

\vspace{-3mm}

\begin{figure}[H]
    \centering
    \includegraphics[width=\linewidth]{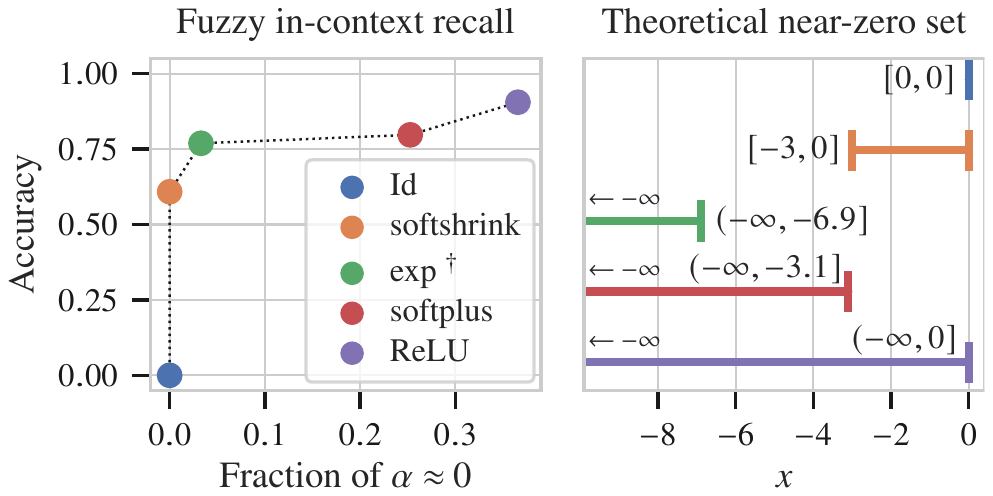}
    \caption{\footnotesize{The wider a readout's near-zero region (right), the more robustly a model can forget irrelevant tokens, which translates directly into higher recall accuracy (left)~\cite{sieber2025design}.}}
    \label{fig:zero-level-set}
\end{figure}

\vspace{-3mm}

All in all, stability, memory horizon, selectivity, and state dimension thus become explicit \emph{design variables}, and the token mixer becomes an object on which stability analysis, eigenvalue placement, and the other standard tools of control apply directly.

\end{sidebar}

\subsection{\four{In the Layers}}

Operating at the depth axis, each layer block implements the update rule $\fivem{\mathcal{B}}[\onem{f}^{(\ell)}]:\mathbb{R}^{T\times d}\!\to\!\mathbb{R}^{T\times d}$.

\paragraph{\four{The dynamics in the depth axis}}

We identify the set of sequential layers as a discrete-time, finite-horizon dynamical system along the depth axis: 
\begin{equation}
    \label{eq:layers-dynamics}
    h_\ell = \fivem{\mathcal{B}}[\onem{f}^{(\ell)}](h_{\ell-1}),
\end{equation}
where the \emph{state} is the activation $h_\ell \in \mathbb{R}^{T\times d}$, the \emph{transition map} is the scaffolded block $\fivem{\mathcal{B}}[\onem{f}^{(\ell)}]$, and the \emph{horizon} is the network depth $L$. At inference time, with the weights frozen, this system runs \emph{autonomously}: once the prompt $s$ is embedded into $h_0 = E(s)$, the activation evolves forward with no exogenous input, until the final state $h_L$ is reached. 

Studying this autonomous flow is itself of considerable interest: the trajectory $h_0\to h_1\to\cdots\to h_L$ is precisely the object that mechanistic interpretability seeks to explain: how internal representations are progressively transformed into behavior. A dynamical-systems reading invites the usual questions of where the flow concentrates, what its fixed points are, and which states it can reach. In practice, however, this analysis is largely out of reach: the transition maps $\fivem{\mathcal{B}}[\onem{f}^{(\ell)}]$ are very high-dimensional and strongly nonlinear through the nonlinearities of the token mixer and channel function, so the autonomous dynamics admit none of the tools available for linear systems. We can rarely predict or certify the behavior of \eqref{eq:layers-dynamics} by inspection alone.

\begin{sidebarfig}[AppleBlue5!18!white]
    \includegraphics[width=\linewidth]{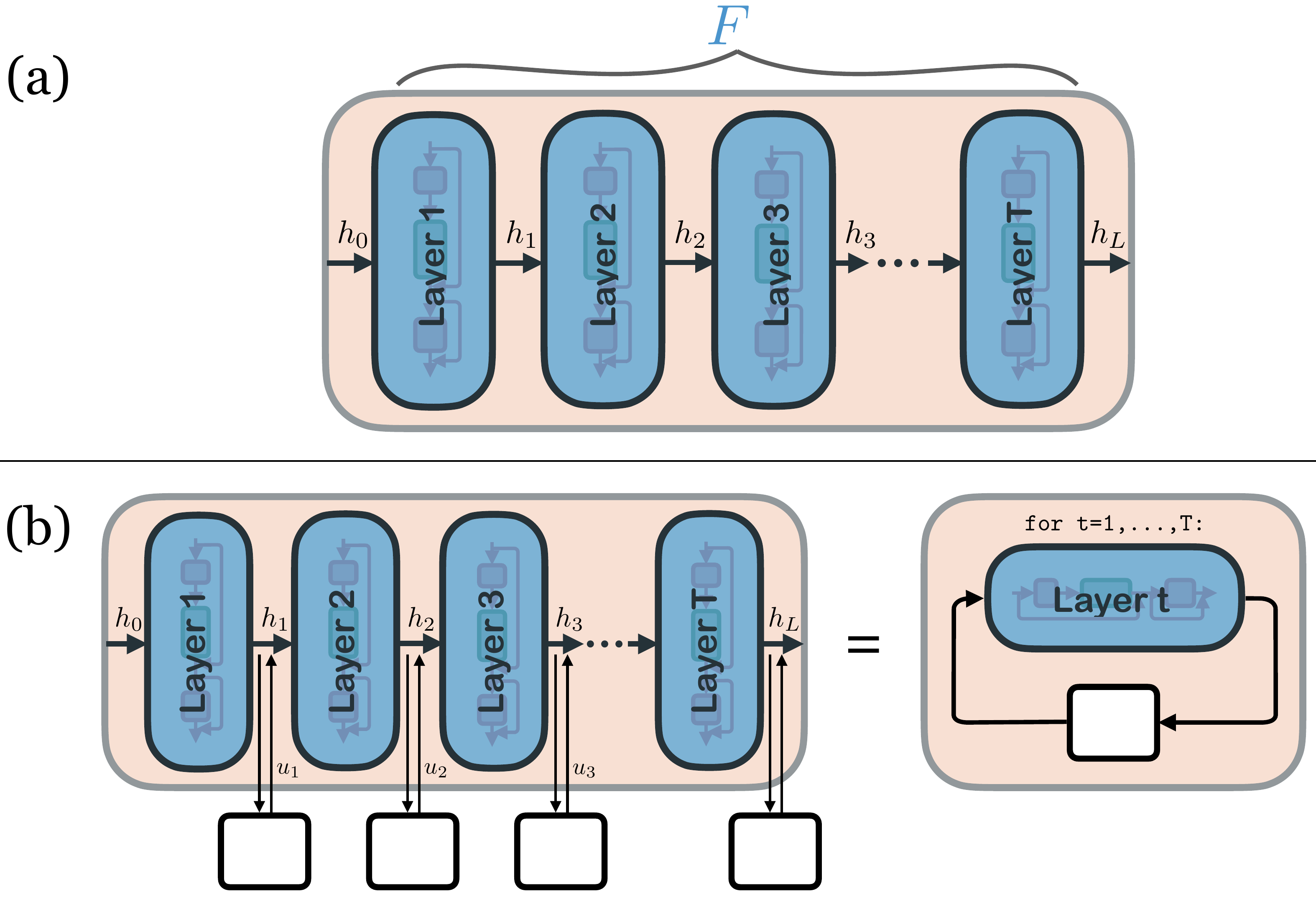}
    \caption{\footnotesize{(a) Autonomous dynamics~\eqref{eq:layers-dynamics}: the activation $h_0,\dots,h_L$ propagates through $L$ scaffolded layers with no exogenous input. (b) Controlled dynamics~\eqref{eq:layers-control}: a depth-axis control input $u_\ell$ is injected before each layer (left), equivalently written as a control loop applied at every layer $t=1,\dots,L$ (right). The control-input boxes (white) mark where each $u_t$ enters the layer stack.}}
    \label{fig:layers-combined}
\end{sidebarfig}

\paragraph{\four{Controlling the dynamics through depth}}
While characterizing the dynamics along the depth axis can result in a nonlinear high-dimensional problem, this is exactly the regime in which a control-inspired approach is most valuable. Rather than fully characterizing the autonomous dynamics, we shape its trajectory online by 

\noindent
introducing a control input into \eqref{eq:layers-dynamics}:
\begin{equation}
  \label{eq:layers-control}
    h_0 = E(s), \quad \tilde h_\ell = h_\ell + u_\ell(h_\ell), \quad
    h_{\ell+1} = \fivem{\mathcal{B}}\!\left[\onem{f}^{(\ell+1)}\right]\!(\tilde h_\ell),
\end{equation}
with $\ell = 0,\dots,L-1$, where $u_\ell(h_\ell)\in\mathbb{R}^{T\times d}$ is a depth-axis control input. Since $u_\ell$ is a function of the current activation, \eqref{eq:layers-control} is a \emph{feedback} law rather than a fixed open-loop perturbation. This idea is powerful: whereas most approaches modify activation trajectories by training (i.e., by altering the weights, and with them the autonomous dynamics \eqref{eq:layers-dynamics}), introducing a control input addresses the same goal in a far more transparent, efficient, and adaptable manner. Because it acts on activations rather than weights, it requires no retraining: the controller is a thin wrapper around each layer, agnostic to internal computations.

In order to design an input $u$ that successfully modifies the behavior of the system \eqref{eq:layers-control}, two properties are required. First, we need a way to \emph{sense} the behaviorally relevant quantity we wish to influence, which we refer to as a \emph{feature} $\zeta\in\mathbb{R}^n$ (examples of a feature include toxicity or sentiment in a language model, or the speed or gripper state of a robot in an embodied one). This sensing or observation has to occur directly from the activation, since the feature is not exposed in the state but latent within it. Second, we need a way to \emph{actuate} on the state so as to drive that feature toward a desired region. Hence, the ``sensor" turns the high-dimensional activation into an interpretable reading of the feature; the ``actuator" turns a desired change in that reading into a perturbation of the activation. This idea is widely present across the \emph{activation steering}~\cite{rimsky2023caa} literature; yet few works exploit the evolution through depth and instead focus on sensing and actuating in only a few layers. It is this idea of unrolled feedback that affords very simple sensing and actuation strategies that result in good performance. Sidebar \emph{Optimal Interventions to Control Layer Dynamics} works out one such sensor--actuator pair in closed form.

\begin{sidebar}[AppleBlue5!18!white]{Optimal Interventions to Control Layer Dynamics~\cite{liseco,buurmeijer2026vla}}

The loop described above is agnostic to the form of the sensor and the actuator; we now give one concrete, lightweight realization that is fully closed-form and adds negligible inference cost. This construction has  been used to steer the behavior of LLMs~\cite{liseco} as well as embodied VLA policies~\cite{buurmeijer2026vla}, illustrating that the loop transfers across modalities essentially unchanged.

For the sensor, the linear representation hypothesis~\cite{lrh} states that features are approximately linearly decodable from intermediate activations. This leads to an affine sensor:
\begin{equation}
  \label{eq:sensor}
  \zeta_\ell = \mathcal{S}_\ell(h_\ell) = W_\ell^{\top} h_\ell + b_\ell,
\end{equation}
whose parameters $(W_\ell,b_\ell)$ are fit offline by regressing activations onto feature labels. 
A user-specified band  $\mathcal{D}=[\zeta_{\min},\zeta_{\max}]$ on a scalar feature then carves out, through the sensor, an \emph{allowed region} of state space: $\{h : \zeta_{\min}\le W_\ell^{\top} h + b_\ell \le \zeta_{\max}\}$.

For the actuator, the \emph{smallest} intervention that brings the feature inside the band is chosen, so that the model's natural trajectory is disturbed as little as possible. Solved independently at each controlled depth (a per-layer relaxation of the horizon-wide control problem), this reads:
\begin{equation}
  \label{eq:actuator-opt}
  u_\ell = \arg\min_{u\in\mathbb{R}^{T\times d}} \ \|u\|_2^2
  \quad\text{s.t.}\quad \mathcal{S}_\ell(h_\ell + u)\in\mathcal{D}.
\end{equation}
With the linear sensor \eqref{eq:sensor}, problem \eqref{eq:actuator-opt} is the Euclidean projection of $h_\ell$ onto the slab and admits the closed-form
\begin{equation}
  \label{eq:liseco-law}
  u_\ell =
  \begin{cases}
    (\zeta_{\max}-\zeta_\ell)\,\dfrac{W_\ell}{\|W_\ell\|_2^2}, & \zeta_\ell > \zeta_{\max},\\[2.0ex]
    (\zeta_{\min}-\zeta_\ell)\,\dfrac{W_\ell}{\|W_\ell\|_2^2}, & \zeta_\ell < \zeta_{\min},\\[2.0ex]
    0, & \text{otherwise,}
  \end{cases}
\end{equation}
where $\zeta_\ell=\mathcal{S}_\ell(h_\ell)$ is the current reading~\cite{liseco}. Notice that when the feature is already in range the actuation vanishes, $u_\ell = 0$: the law has a dead zone and is therefore minimally invasive, unlike fixed-magnitude steering vectors that perturb the stream regardless of need. When the feature leaves the band, the correction is the shortest vector returning it to the violated boundary (directed along the sensor's normal $W_\ell$ and scaled by how far the reading lies outside $\mathcal{D}$), and both its direction and its magnitude depend on the current state. 

\emph{Example of results (directly extracted from \cite{liseco,buurmeijer2026vla}):} Because the actuator only nudges the state exactly as much as needed to satisfy the constraint, this controlled behavior tends to leave everything else about the model's output close to its uncontrolled baseline while satisfying the constraint. That is precisely the trade-off that matters in practice: how much does enforcing the constraint cost elsewhere? Empirically, remarkably little, both for LLM generation and for a robot policy.

\begin{figure}[H]
    \centering
    \begin{minipage}{0.48\linewidth}
        \centering
        \includegraphics[width=\linewidth]{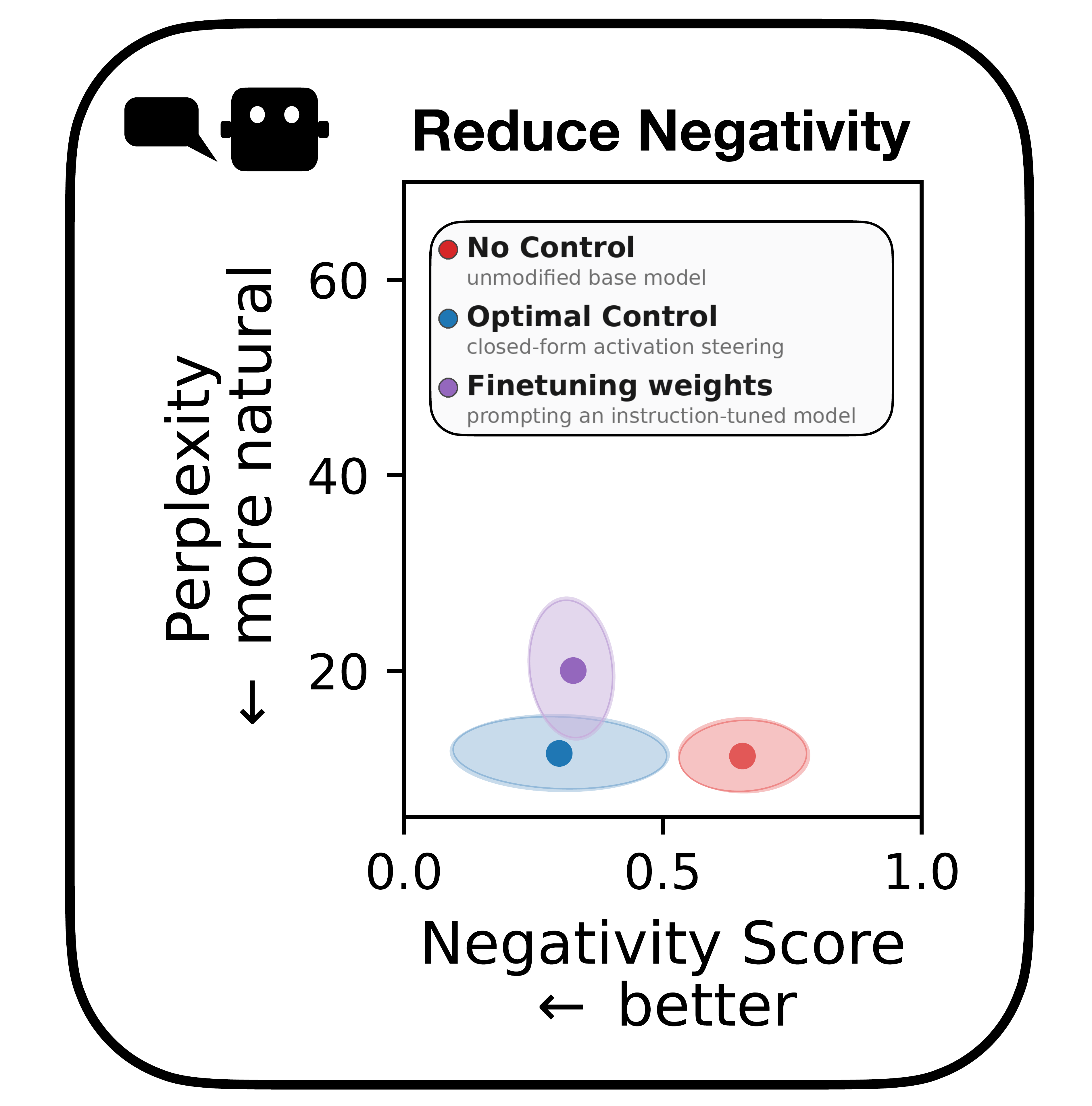}
    \end{minipage}
    \hfill
    \begin{minipage}{0.48\linewidth}
        \centering
        \includegraphics[width=\linewidth]{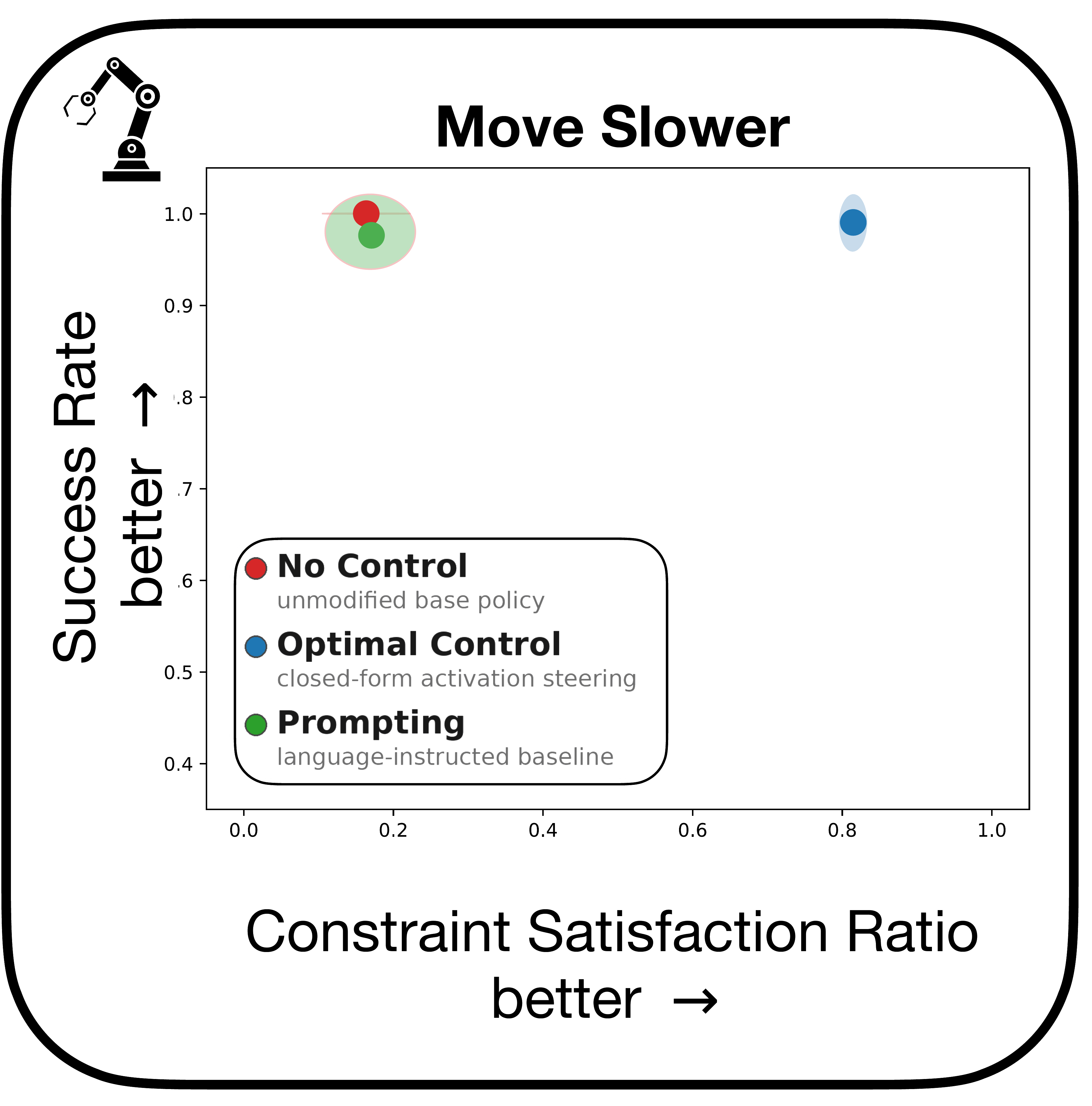}
    \end{minipage}
    \caption{\footnotesize{(a) On Llama-3-8B for language on a sentiment benchmark, the intervention reduces negativity while maintaining the naturalness of the uncontrolled baseline~\cite{liseco}. (b) On a $\pi_{0.5}$ robot policy in the Libero simulator, the intervention achieves slower speed while maintaining the task success rate of the uncontrolled baseline~\cite{buurmeijer2026vla}.}}
    \label{fig:optimal-control-negativity}
\end{figure}

All in all, casting the sensor--actuator loop as a per-layer intervention turns feature control into a closed-form, training-free primitive. The natural next step is to move beyond this per-layer relaxation~\eqref{eq:actuator-opt} toward coordinating interventions across the whole depth axis, so that the correction budget is spent where it matters most over the horizon rather than greedily at each layer; a second is extending the affine sensor~\eqref{eq:sensor} to richer, non-linear probes for a wider set of features.

\end{sidebar}

\subsection{\three{In the Deep Architecture}}
Operating at the generation axis, the deep architecture places the layer stack in a feedback loop and iterates it until a sample is complete, as captured in equation~\eqref{eq:generation}. 

\paragraph{\three{The dynamics in the generation axis}}

The model dynamics along the generation axis are captured by a discrete-time, finite-horizon dynamical system of the form:
\begin{equation}\label{eq:autonomous-generation}
  s_{k+1} = \threem{\phi}\!\left(s_k,\, \fourm{F_\theta}(s_k)\right),\qquad
  k = 0,\dots,K-1,\qquad s_0 \sim p_0,
\end{equation}
where $s_K$ is the state, $\threem{\phi}$ the (stochastic) decoding rule, and $K$ is the generation horizon. This axis is nested: advancing $s_k$ a single step requires running the entire depth  axis to completion, since each call to $\threem{\phi}$ evaluates the backbone $\fourm{F_\theta}$.

Four familiar decoding rules make this abstract loop concrete, each instantiating $\threem{\phi}$ differently while leaving~\eqref{eq:autonomous-generation} itself untouched -- each an autonomous special case of~\eqref{eq:autonomous-generation}, differing only in how $\threem{\phi}$ turns $\fourm{F_\theta}(s_k)$ into $s_{k+1}$:

\begin{itemize}
\item \textbf{Autoregressive}: the state $s_k\in\mathbb{R}^{(T_0+k)\times d}$ grows by one sampled token per step, starting from the prompt:
\[
\begin{aligned}
x_{k+1}&\sim\mathrm{Categorical}\big(\mathrm{softmax}(\fourm{F_\theta}(s_k))\big), \\
s_{k+1}&=[\,s_k\,;\,x_{k+1}\,].
\end{aligned}
\]
\item \textbf{Masked / iterative}: the state $s_k\in\mathbb{R}^{T\times d}$ stays fixed length and fills in a growing subset of positions, starting all-masked, with $M_k$ the set of still-masked
positions and $U_k \subset M_k$ those unmasked at step $k$:
\[
\begin{aligned}
(s_{k+1})_i &= x_i\sim\mathrm{Categorical}\big(\mathrm{softmax}(\fourm{F_\theta}(s_k))_i\big), \\
&\text{for } i\in U_k\subseteq M_k, \text{ else } (s_{k+1})_i=(s_k)_i.
\end{aligned}
\]
\item \textbf{Diffusion}: the state $s_k\in\mathbb{R}^{T\times d}$ stays fixed length and is refined by a stochastic denoising update, starting from $\mathcal{N}(0,I)$:
\[
\begin{aligned}
s_{k+1}&=\dfrac{1}{\sqrt{\alpha_k}}\Big(s_k-\dfrac{1-\alpha_k}{\sqrt{1-\bar\alpha_k}}\,\fourm{F_\theta}(s_k)\Big) \\
&\quad+\sigma_k z_k, \quad z_k\sim\mathcal{N}(0,I),
\end{aligned}
\]
where $\beta_k\in(0,1)$ is the forward noise-variance schedule, $\alpha_k:=1-\beta_k$ is the fraction of signal retained at step $k$, and $\bar\alpha_k:=\prod_{j\le k}\alpha_j$ is its cumulative product; $\sigma_k$ is the fixed noise scale re-injected at each reverse step. 
\item \textbf{Flow matching}: the state $s_k\in\mathbb{R}^{T\times d}$ stays fixed length and is refined by a deterministic step along a learned velocity field, starting from $\mathcal{N}(0,I)$:
\[
s_{k+1} = s_k + \tfrac{1}{K}\,\fourm{F_\theta}(s_k).
\]
\end{itemize}

\begin{sidebarfig}[AppleOrange5!18!white]
    \includegraphics[width=.8\linewidth]{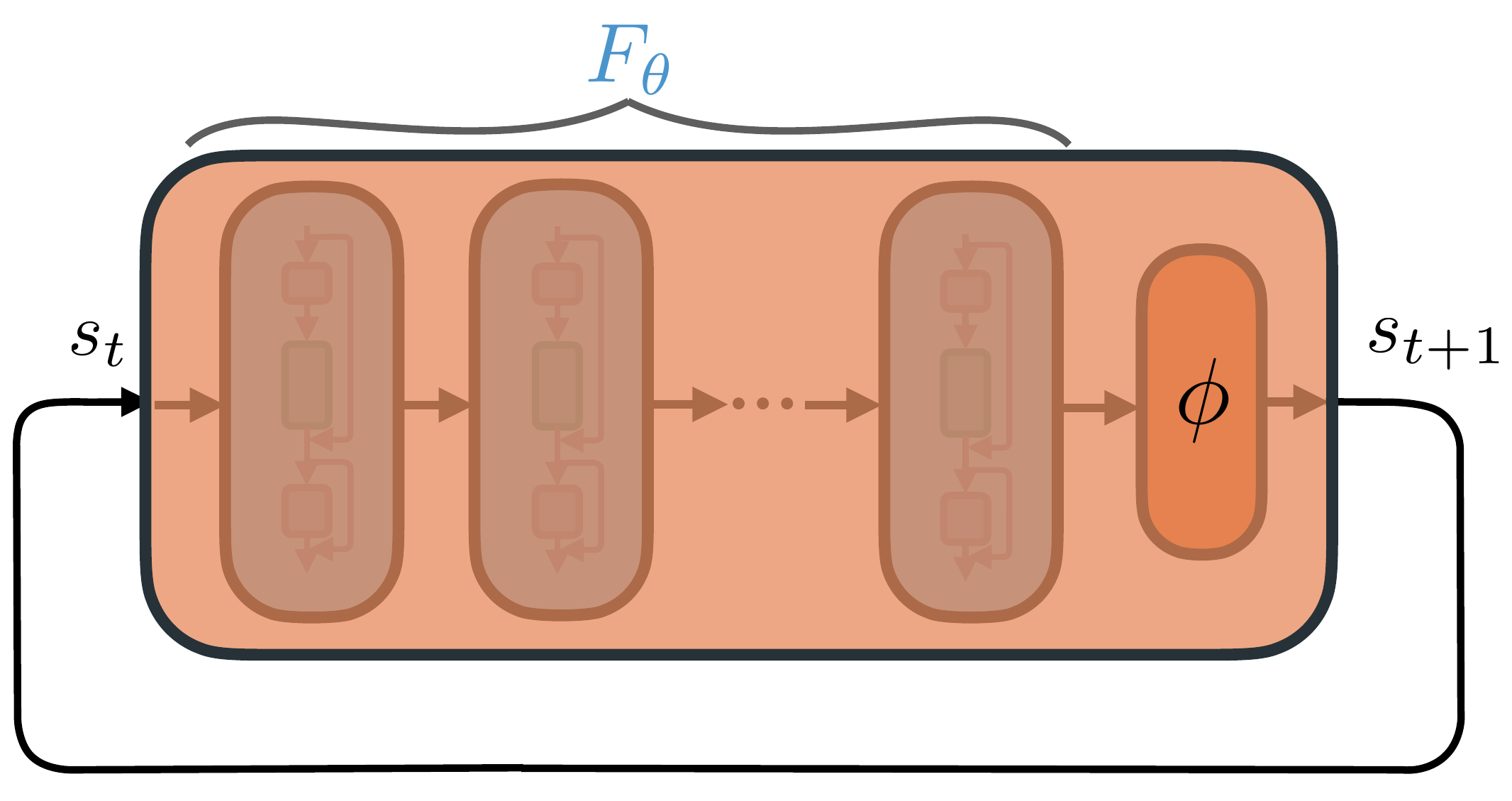}
    \caption{\footnotesize{The generation loop along the generation axis: the autonomous dynamics of~\eqref{eq:autonomous-generation}, iterating the backbone $\fourm{F_\theta}$ through the decoding rule $\threem{\phi}$ (outer loop) for $K$ steps with no exogenous input. Each iteration nests a full pass through the depth axis (inner loop), and, within it, the token axis.}}
    \label{fig:generation-dynamics}
\end{sidebarfig}

\paragraph{\three{Controlling the dynamics through generation}}
Formally, the control input enters the loop as
\begin{equation}\label{eq:controlled-generation}
  \begin{aligned}
    s_{k+1} &= \threem{\phi}\!\left(s_k,\, \fourm{F_\theta}(s_k),\, u_k\right), \ u_k &= u^\star(s_k,k),
  \end{aligned}
\end{equation}
with $k=0,\dots,K-1$, where $u_k$ is a generation-axis control input shaping how the decoding rule advances the state at each step. Since $u_k$ is a function of the current state, \eqref{eq:controlled-generation} is a \emph{feedback} law rather than a fixed open-loop perturbation, exactly as in the depth-axis case above.

In order to design an input $u$ that successfully shapes the trajectory $s_0,\dots,s_K$ toward a target (a data distribution, a desired attribute, a partially specified sequence), the same two ingredients as before are required. First, a way to \emph{sense} the behaviorally relevant quantity we wish to influence -- a feature of the current, possibly partial, generation, such as its toxicity or sentiment, or how closely it matches a target attribute. Second, a way to \emph{actuate} on the state so as to drive that feature toward the target over the remaining $K-k$ steps, by injecting $u_k$ into the decoding rule at every generation step. Unlike the depth axis, where each layer's correction only has to account for the remaining depth of a single forward pass, the generation-axis controller must reason over the entire remaining trajectory: every choice of $u_k$ changes the state that all later decoding steps, and hence all later corrections, will act on.

\begin{sidebar}[AppleOrange5!18!white]{Optimal Control for the Generation Loop~\cite{optimalcontrolgen}}

The work in \cite{optimalcontrolgen} proposes to cast generation as a stochastic optimal control problem, where the optimal controller $u^\star$ is the one minimizing a cost that trades a terminal cost (fidelity of the final sample $s_K$ to the target) against a running cost penalizing the trajectory's transport effort $\sum_k\|u_k\|^2$. The corresponding value function $V(s,k)$ (the minimal expected cost-to-go) satisfies a Hamilton--Jacobi--Bellman equation whose solution yields the closed-loop law
\begin{equation}\label{eq:hjb-generation}
  \begin{aligned}
    -\partial_k V(s,k) &= \min_{u} \, H\big(s,u,\nabla_s V\big), \\
    H(s,u,p) &= \tfrac12\|u\|^2 + u\cdot p, \\
    u^\star(s,k) &= -\nabla_s V(s,k).
  \end{aligned}
\end{equation}
Unlike the per-layer sensor--actuator law of~\eqref{eq:liseco-law}, the controller here is not restricted to a small, interpretable correction: it is a full closed-loop policy shaping the entire generation trajectory toward the goal.

Solving~\eqref{eq:hjb-generation} exactly is, in general, intractable: $V$ is a value function over the full, high-dimensional state space of the generation axis, and no closed form exists for real backbones $\fourm{F_\theta}$. \cite{optimalcontrolgen} sidesteps solving the PDE directly by exploiting a duality between this Eulerian (PDE) formulation and a Lagrangian (path-based) one: rather than solving for $V$, they learn the optimal vector field $u^\star$ directly via flow matching, over a \emph{rectified latent control space} obtained by training a variational autoencoder whose latent geometry is smooth and well-conditioned (in contrast to the raw token/embedding space, where $\nabla_s V$ is ill-defined or ill-conditioned). The resulting model, Manta-LM, realizes a genuinely closed-loop, full-horizon controller: at every step it senses the current state and corrects it along the learned descent direction of $V$, rather than committing once (as in autoregressive decoding) or correcting blindly (as in diffusion over a poorly conditioned space).

\emph{Example of results (reproduced from Fig.~2(d) of~\cite{optimalcontrolgen}):} Tracking the mean $L_2$ error against how far generation has progressed, autoregressive decoding's error climbs early and then plateaus at a high level -- exactly the irreversible error compounding of an impulsive, open-loop correction discussed above -- while discrete diffusion's error falls only gradually and remains high through most of the trajectory. The closed-loop optimal-control policy instead drives the error down fastest and furthest, ending roughly an order of magnitude below either baseline by the end of generation: direct evidence that a full-horizon controller, not just a better per-step correction, is what closes the gap.

\begin{figure}[H]
    \centering
    \includegraphics[width=0.85\linewidth]{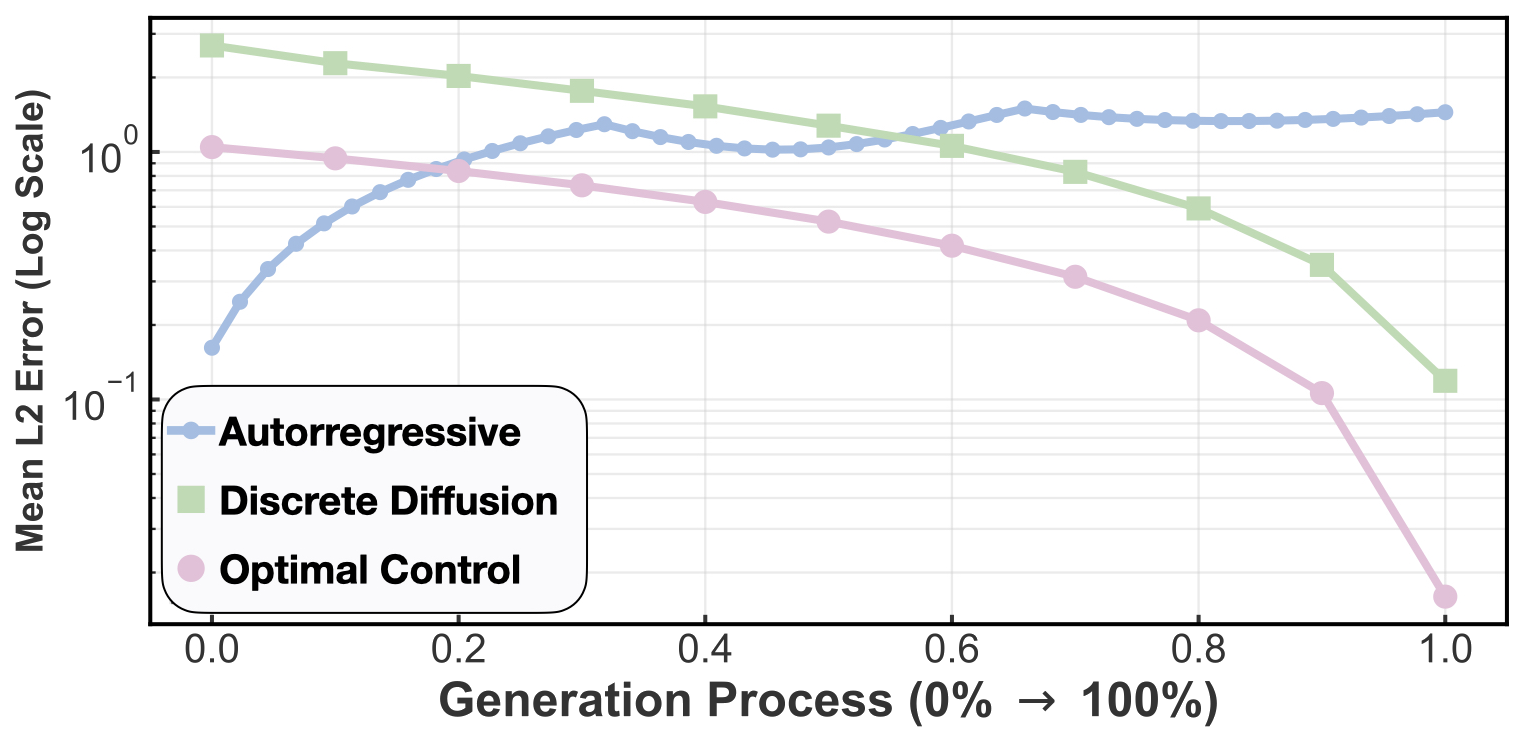}
    \caption{\footnotesize{Mean $L_2$ error (log scale) as generation progresses from 0\% to 100\% complete, for autoregressive decoding, discrete diffusion, and the closed-loop optimal-control controller. Autoregressive error compounds and plateaus early; diffusion's error falls only gradually; the closed-loop controller drives error down fastest and furthest; reproduced from Fig.~2(d) of~\cite{optimalcontrolgen}.}}
    \label{fig:generation-l2error}
\end{figure}

All in all, casting generation as a stochastic optimal control problem turns the choice of decoding rule from an architectural commitment into a design variable: approximating the full closed-loop law $u^\star$ directly is what closes the efficiency-fidelity gap left open by the autonomous rules above. The natural next step is to move beyond a fixed terminal cost toward goals specified at inference time (e.g.\ a user-supplied constraint or reward), turning the generation-axis controller into the same kind of flexible, training-free primitive that the depth-axis sensor--actuator law of~\eqref{eq:liseco-law} provides one level down.

\end{sidebar}

This full-horizon character is what separates a genuine controller from the autonomous rules above, of which we take autoregressive and diffusion decoding as the two representative cases. Autoregressive decoding is an \emph{impulsive, open-loop} controller: one greedy, local correction per token, with no feedback on how the full trajectory turns out, so early errors compound irreversibly. Diffusion decoding applies a dense correction at every step, but either over a discrete vocabulary with no usable gradient, or over a raw embedding space whose geometry ill-conditions the gradient that does exist. Both are special cases of the general control problem \eqref{eq:autonomous-generation}, recovered from \eqref{eq:controlled-generation} by a fixed choice of $u_k$. Sidebar \emph{Optimal Control for the Generation Loop} works out the full closed-loop controller that these two rules approximate.

\subsection{\two{In AI-World Interactions}}
Operating at the interaction axis, the model is no longer a function we evaluate but a policy acting in closed loop with the world, as captured in  equation~\eqref{eq:interaction}. 

\paragraph{\two{The dynamics in the interaction axis}}
The model is seen as an agent that at time-step $n$ produces an action $a_n$ based on context $c_n$. This action impacts the environment, and the resulting effects are folded back into the context $c_{n+1}$, over a horizon of $N$ exchanges. Two features set this level apart from every level below it. First, advancing the loop a single step requires running an \emph{entire} generation to completion: the interaction axis sits at the slowest time scale, with the generation, depth, and token axes nested inside each of its steps. Second, and more consequentially for control, the loop is closed through the external world rather than inside the model, which introduces a level of complexity and uncertainty that calls for fully data-driven methods instead of relying on analytical forms.

We identify the closed loop between model and world as a discrete-time, finite-horizon dynamical system along the interaction axis as captured in equation \eqref{eq:interaction} over a sequence of exchanges
$n = 0,\dots,N-1$ with \emph{horizon} N, given an initial context $c_0$, with the \emph{state} being the context $c_n$, and the \emph{transition map} as the interaction rule $\twom{\psi}$ composed with the generation-axis policy $\threem{\Phi}$ that produces the action. 
No controller is present yet: the model reacts to whatever the world returns, and the world reacts to whatever the model does, with no external agent steering either side toward a goal.

\paragraph{\two{Controlling the dynamics through interactions}}

We take a goal-oriented perspective, where we assume that the interaction with the world is intended to control the behavior of the system. Hence, the closed-loop AI-world interaction along the interaction axis can be modeled as a control process, where in this case a prompt-injector (could be a human user or another AI agent) is seen as the controller whose goal is for the AI system to reach a specific output. In this case, the controlled dynamics take the form: 
\begin{equation}\label{eq:controlled-interaction}
  a_n = \threem{\Phi}(c_n), \qquad
  c_{n+1} = \twom{\psi}\!\left(c_n, a_n, \iota_n\right), \qquad
  \zeta_n = \twom{\Gamma}(a_n),
\end{equation}
where $c_0$ is the initial context (the initial prompt); $\iota_n \in \mathcal{I}$  is the intervention the controller injects at step $n$, drawn from an intervention space $\mathcal{I}$ (a prompt, an in-context example, or any other admissible degree of freedom); and $\twom\Gamma$ is a readout map. This readout map acts as an interaction-level \emph{sensor}, extracting from each action the behavioral feature $\zeta_n \in \mathcal{Z}$ the controller cares about. The interaction-level \emph{actuation} happens through  $\twom\psi$, which is itself unmodeled: the
controller sets $\iota_n$ but cannot predict how the world will fold it into $c_{n+1}$. 
We note that the dynamics in the interaction axis \eqref{eq:controlled-interaction} are fundamentally harder to govern than the ones in the inner axes: the controller acts \emph{only} through $\iota_n$ and observes \emph{only} $\zeta_n$; it can neither read nor write the internal state carried by $\psi$. Equation \eqref{eq:controlled-interaction} is therefore an output-feedback loop, in contrast to the white-box state feedback of the token and depth axes, where it is possible to sense and actuate the states directly.

\section{Challenges and Opportunities}

This article set out from a simple premise: current generative models are capable but come with no principled design or guarantees on their behavior; yet similar to what happened in other engineering disciplines, we argue that control theory has a big role to play to create principled and controllable AI architectures. In this article, we showed

\begin{sidebarfig}[ApplePurple5!18!white]
    \begin{subfigure}[t]{0.54\linewidth}
        \centering
        \includegraphics[width=\linewidth]{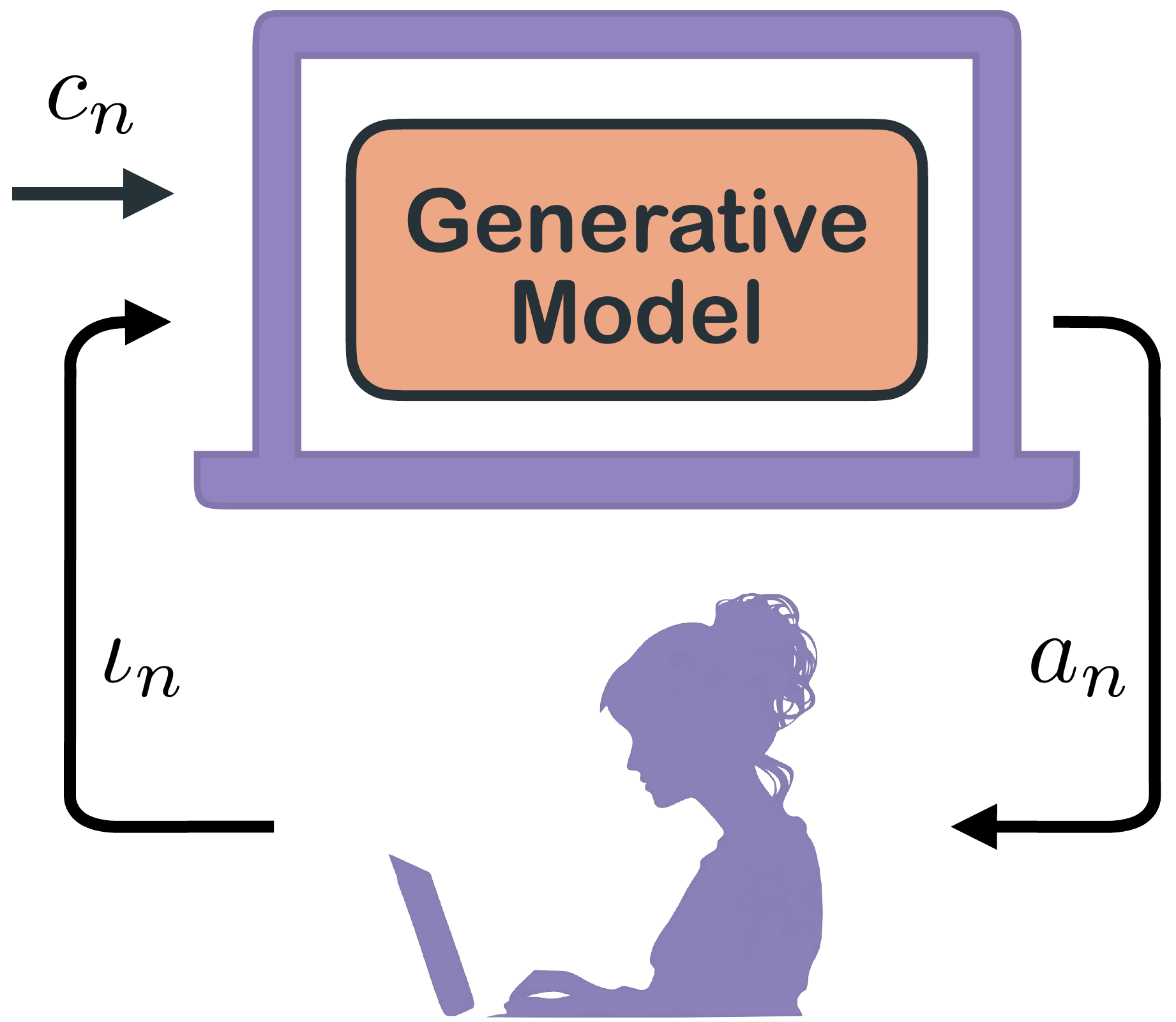}
        \caption{}
        \label{fig:interaction-illustration}
    \end{subfigure}
    \hfill
    \begin{subfigure}[t]{0.44\linewidth}
        \centering
        \vspace{-35mm}
        \resizebox{!}{1.02\linewidth}{%
        \begin{tikzpicture}[
          every node/.style={font=\fontsize{13}{15}\selectfont},
          cbox/.style={rectangle, rounded corners, draw=black!45, fill=black!4, text width=6.6cm, align=left, inner sep=5pt},
          abox/.style={rectangle, rounded corners, draw=AppleOrange5, fill=AppleOrange5!10!white, text width=6.6cm, align=left, inner sep=5pt},
          ibox/.style={rectangle, rounded corners, draw=ApplePurple5, fill=ApplePurple5!10!white, text width=6.6cm, align=left, inner sep=5pt},
          arr/.style={-{Latex[length=2mm]}, thick, black!55}
        ]
          \node[cbox] (c0) {$c_0=$ \textit{``Hello''}};
          \node[abox, below=5pt of c0] (a0) {$a_0=\threem{\Phi}(c_0)=$ \textit{``How can I help you today?''}};
          \node[ibox, below=5pt of a0] (i0) {$\iota_0=$ \textit{``Give me a word with four letters''}};
          \node[cbox, below=5pt of i0] (c1) {$c_1=\twom{\psi}(c_0,a_0,\iota_0)=c_0+a_0+\iota_0$};
          \node[abox, below=5pt of c1] (a1) {$a_1=\threem{\Phi}(c_1)=$ \textit{``Apple''}};
          \node[ibox, below=5pt of a1] (i1) {$\iota_1=$ \textit{``No, that has five letters''}};
          \node[cbox, below=5pt of i1] (c2) {$c_2=\twom{\psi}(c_1,a_1,\iota_1)=c_1+a_1+\iota_1$};
          \node[abox, below=5pt of c2] (a2) {$a_2=\threem{\Phi}(c_2)=$ \textit{``Kiwi''}};
          \draw[arr] (c0) -- (a0);
          \draw[arr] (a0) -- (i0);
          \draw[arr] (i0) -- (c1);
          \draw[arr] (c1) -- (a1);
          \draw[arr] (a1) -- (i1);
          \draw[arr] (i1) -- (c2);
          \draw[arr] (c2) -- (a2);
        \end{tikzpicture}%
        }
        \caption{}
        \label{fig:interaction-example}
    \end{subfigure}
    \vspace{-2pt}
    \caption{\footnotesize{The controlled interaction dynamics~\eqref{eq:controlled-interaction}. (a) The model produces an action $a_n=\threem{\Phi}(c_n)$ from the current context $c_n$; the controller (here, a human user) observes the action and injects an intervention $\iota_n$, and both are folded into the next context $c_{n+1}=\twom{\psi}(c_n,a_n,\iota_n)$. (b) A worked instance: the controller repeatedly injects an intervention $\iota_n$ (purple) that steers the model's next action $a_n$ (orange), while the context $c_n$ (grey) accumulates the full exchange, $c_{n+1}=c_n+a_n+\iota_n$, where $+$ denotes string concatenation.}}
    \label{fig:interaction-combined}
\end{sidebarfig}

\begin{sidebar}[ApplePurple5!18!white]{What Outcomes can be Reached by Controlling Interaction Dynamics?~\cite{genctrl}}

Choosing \emph{how} to intervene is a rich question on its own, but it presupposes a more basic one: is the desired outcome  attainable at all? This is the question of \emph{reachability}.

\medskip\noindent\emph{\textbf{Reachability with a fixed context.}}\; Given an initial context $c_0$, the reachable set collects every feature value produced by some admissible intervention sequence over $n$ exchanges,
\begin{equation}\label{eq:reach-set}
  \mathcal{R}(c_0, \mathcal{I}, n)
   = \bigl\{\, \tilde\zeta \in \mathcal{Z} \;\big|\;
     \exists\, \iota_0,\dots,\iota_{n-1}\in\mathcal{I}
     \text{ with } \zeta_n = \tilde\zeta \,\bigr\},
\end{equation}
so the target is attainable from $c_0$ only if $\mathcal{D} \subseteq \mathcal{R}(c_0, \mathcal{I}, n)$.

Because $\twom \psi$ and $\threem \Phi$ are unknown, nonlinear, and high-dimensional,
$\mathcal{R}$ admits no analytic form and must be estimated from rollouts of the loop itself. Two facts shape the estimator. First, actions are drawn from a finite vocabulary, so the reachable feature set is \emph{countable}: a continuous estimate would be vacuously large, and we instead coarse-grain any continuous $\mathcal{Z}$ into cells of resolution $\kappa$, yielding $Q$ cells in total. Second, sampling admits only a probabilistic guarantee. Drawing $J$ independent intervention rollouts and reading out their features gives the Monte~Carlo estimate $\hat{\mathcal{R}}(c_0, \mathcal{I}, n)$ as the union of the visited cells; this estimate is \emph{distribution-free} and assumes nothing about $\psi$ beyond boundedness of $\mathcal{Z}$. For a failure probability $\delta$ and a precision $\epsilon$, it suffices to draw
\begin{equation}\label{eq:reach-bound}
  J \;\geq\; \max\!\left(\,Q,\;
    \frac{\log(\delta/Q)}{\log(1-\epsilon)}\,\right)
\end{equation}
rollouts to cover every cell with probability at least $1-\delta$. The estimate is then actionable: if $\mathcal{D}\not\subseteq\hat{\mathcal{R}}(c_0,\mathcal{I},n)$, the desired behavior is unreachable from $c_0$ with probability at least $1-\delta$.

\medskip\noindent\emph{\textbf{Reachability from any context.}}\; The set $\mathcal{R}(c_0,\mathcal{I},n)$ certifies what a \emph{single} starting prompt affords; a deployed system, however, must reach its target no matter how the conversation begins. Intersecting the reachable sets over all initial contexts, $\mathcal{R}^{\cap}(\mathcal{I},n) \;=\; \bigcap_{c_0}\,\mathcal{R}(c_0,\mathcal{I},n)$,  collects the features attainable from \emph{every} context. This strict intersection is brittle: a single adversarial context empties it. Hence, a relaxation to a  \emph{$\lambda$-robust} reachable set that keeps the features attainable from at least a $1-\lambda$ fraction of contexts is necessary,
\begin{equation}\label{eq:robust-reach}
  \mathcal{R}_{\lambda}(\mathcal{I},n) \;=\;
   \Bigl\{\, \tilde\zeta \in \mathcal{Z} \;\Big|\;
     \Pr_{c_0 \sim \mathcal{P}_c}\!\left[\, \tilde\zeta \in \mathcal{R}(c_0,\mathcal{I},n)\,\right]
     \geq 1-\lambda \,\Bigr\},
\end{equation}
where $\mathcal{P}_c$ is a distribution over initial contexts and $\lambda \in (0,1)$ is a tolerance. Similar to $\mathcal{R}$, the set $\mathcal{R}_{\lambda}$ has no analytic form and is estimated by sampling, now at two levels. Drawing $J_0$ initial contexts from $\mathcal{P}_c$, the per-context estimate $\hat{\mathcal{R}}(c_0^{(j)},\mathcal{I},n)$ of the previous paragraph can be constructed for each, and retain the cells surviving in at least a $1-\lambda$ fraction of them. To guarantee that the result overcounts the true $\mathcal{R}_{\lambda}$ by a feature-measure below $\omega$ with confidence $1-\delta$, it suffices to sample $J_0$ initial contexts:
\begin{equation}\label{eq:ctrl-bound}
  J_0 \;\geq\; \frac{\log(\omega\,\delta)}{\log(1-\lambda)}.
\end{equation}

\emph{Example of results (directly extracted from~\cite{genctrl}):} On a text-formality control task, the estimator above is used to track how a model's $\lambda$-controllable set grows turn by turn over a 5-turn dialogue. With only zero-shot feedback (no in-context examples, just turn-by-turn corrections), the reachable set for Gemma3-4B stalls: after five turns, most of the output range still sits in the grey, provably-\emph{not}-reachable region, i.e., no admissible sequence of interventions can push this model's output there. Because the estimator is distribution-free and architecture-agnostic, the same procedure applies unchanged to any model, feature, or context, and is released as an open-source toolbox for exactly this kind of controllability audit.

\begin{figure}[H]
    \centering
    \includegraphics[width=0.8\linewidth]{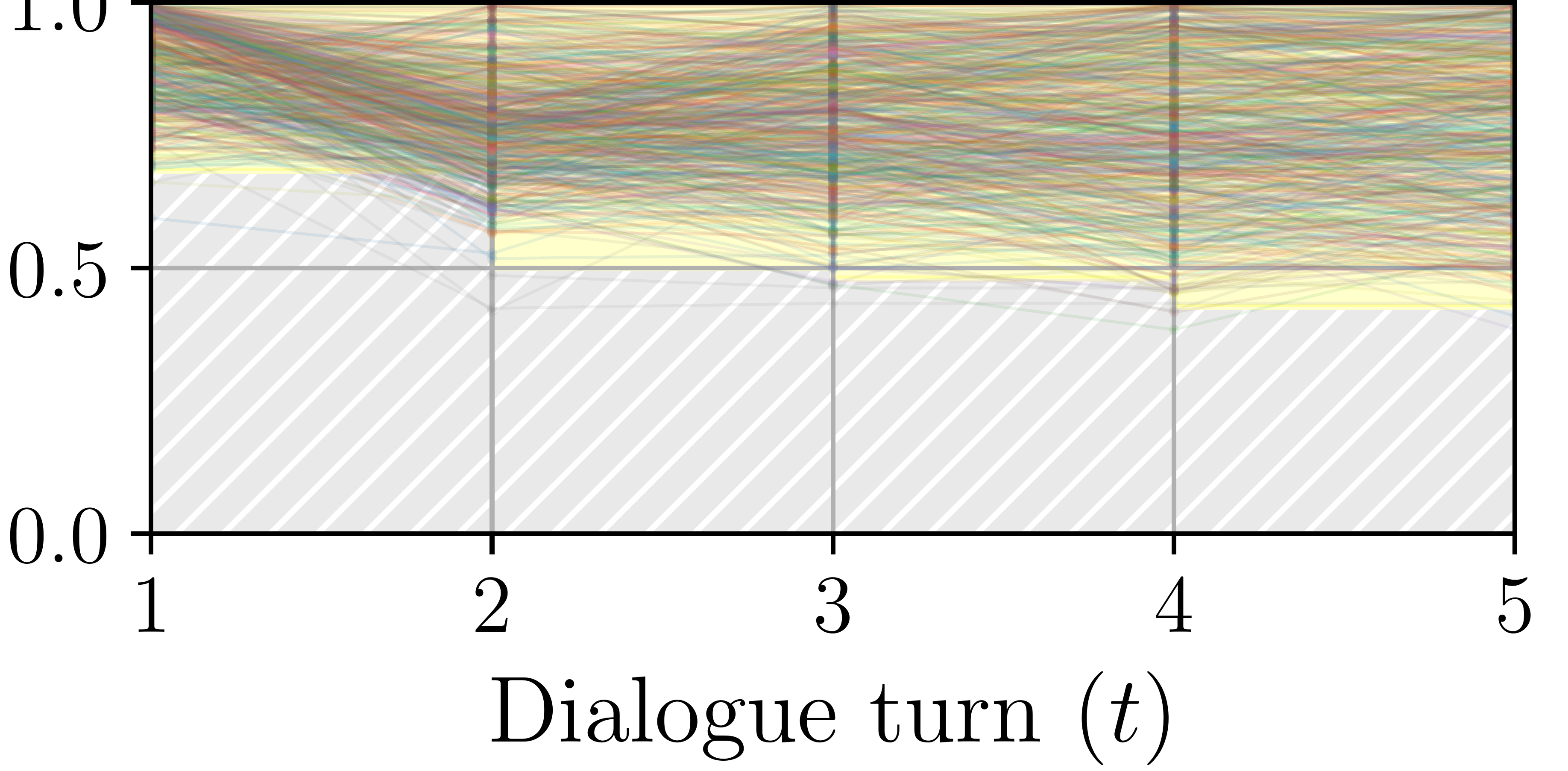}
    \caption{\footnotesize{Gemma3-4B on a text-formality task, zero-shot prompting: each line is a dialogue trajectory from a distinct initial context $c_0$; the yellow band is the estimated $0.1$-controllable set $\hat{\mathcal{R}}(c_0,\mathcal{I},n)$ from~\eqref{eq:reach-bound}, and the grey hatched region is certified \emph{not} reachable. Even after five turns of feedback the controllable set covers barely half the output range, empirically certifying that this model--task--mechanism combination is \emph{not} fully controllable, at confidence $\delta=0.05$~\cite{genctrl}.}}
    \label{fig:genctrl-controllable-set}
\end{figure}

\vspace{-4mm}

All in all, reachability and $\lambda$-robust controllability turn the question of whether a deployed system \emph{can} be steered where a user needs it into something that is certified by sampling rather than assumed. The natural next step is to move from single-feature reachability toward jointly certifying several correlated features at once, and from post-hoc verification toward using the estimated sets to guide the choice of intervention $\iota_n$.

\end{sidebar}

\noindent a comprehensive view on the different dynamics emerging in current generative models, and compiled existing insights that point towards exciting research directions along this path. To do this, we decomposed a generative model into four nested axes along which dynamics occur -- token, depth, generation, and interaction -- each with its own state, its own autonomous dynamics, and its own feedforward and feedback structure. That decomposition helped us look at different components that instantiate such dynamics, and recast a set of existing, previously unrelated-looking techniques -- skip-connection gains, activation steering, scheduled sampling, RLHF's KL penalty -- as instances of a control problem, solved differently at each time scale. The main contribution of this article is not any one of these techniques (all extracted from the existing literature); rather it is the shared framework that lets them be compared, combined, and improved on. As such, the controllable AI framework presented here is a starting point. At nearly every axis, the tools presented in this article are the least powerful ones that already work, which are not necessarily the best suited ones for the underlying control problem. Several of the sidebars above end by naming exactly what is still missing. We collect the most pressing of these below, after the sidebar \emph{What Outcomes can be Reached by Controlling Interaction Dynamics?}, which closes out the interaction axis by asking a question that cuts across all four: not just how to intervene, but whether the desired outcome is reachable at all.

\paragraph{Assembling one controller from four}
Every controller in this article was designed, and evaluated, one axis at a time: the depth-axis law does not know that a generation-axis controller may be running on top of it, and the generation-axis controller does not know that an interaction-axis intervention (prompt) may change the very target it is steering toward. In a real deployment these loops do not run in isolation; they run nested, and a correction made at one time scale shifts the operating point every faster loop nested inside it sees. Classical control has a name and a toolkit for this: cascaded and hierarchical control for time-scale separation. This framework fits exactly this setting, where a slow loop's reference is a fast loop's disturbance. None of it has been brought to bear here yet. Assembling one coordinated controller from four independent ones, rather than four controllers that merely happen to share a state, is the unification that future work in controllable AI aims to achieve.

\paragraph{Architectures designed for control, not retrofitted with it}
Every controller in this article was also built for an architecture that was not designed with control in mind, but rather a correction applied after the fact to a policy trained without it. Yet, as seen in this tutorial, the controller structures that benefit training are very similar to those introduced at inference time. A design discipline in which the feedforward and feedback paths of every axis are control-native from the start does not yet exist. Designing generative models directly as closed loops equipped with controller structures is an important step to create principled and controllable AI architectures by design. 

\paragraph{Scalable, nonlinear, data-driven control}
Throughout this article the question of efficiency is a prominent one, and choices made at every axis reveal this as an important factor both for design and analysis. Most of the approaches introduced rely on very simple control techniques that in turn incur large overhead. The reason behind these choices is that classical control tools were not developed for state spaces of such large dimension: learned value functions, Koopman-operator linearization, model-predictive control with a learned dynamics model, etc. have not yet been systematically brought to bear on a system with billions of parameters. Furthermore, the controller's compute budget can end up competing with the serving cost, latency budget, and memory footprint of the model it is meant to control. Classical control has tackled this problem in some cases, such as \emph{explicit} model predictive control. This consideration is extremely important in controllers designed for AI architectures, and addressing it more generally is an aspect missing in more complex control approaches. Hence, controllable AI presents an exciting opportunity for the field of control theory, as it motivates important questions around scalable techniques, nonlinear controllers, and data-driven methods.

\paragraph{Certification as a policy instrument}
Several of the tools in this article come with a guarantee attached, not just a benchmark score: $\fivem{\lambda}$-skip's condition~\eqref{eq:lambda-condition} provably makes rank collapse arbitrarily slow, the reachability estimator of~\eqref{eq:reach-bound}--\eqref{eq:ctrl-bound} certifies, at a stated confidence level, what a deployed model can and cannot be made to do, etc. That is a different kind of claim than a benchmark score, and arguably an important complementary one as we move towards setting regulatory standards and certifications for AI systems similar to what we have for other engineering systems: a benchmark measures typical-case behavior on a fixed test set that can be overfit or gamed, while a certificate of this kind holds independent of any particular prompt distribution and states plainly the conditions under which it can fail. Extending this style of guarantee -- from a single feature, layer, or axis to a full system, and from a post-hoc audit to a pre-deployment requirement -- is a concrete step to AI evaluation that a regulator or standards body could adopt without waiting for the harder open problems to be solved.

\paragraph{Oversight of increasingly autonomous systems}
The interaction axis is, by this article's own account, the hardest of the four to govern: the controller acts only through an intervention $\iota_n$ and senses only a readout $\zeta_n$, never the internal state carried by $\twom{\psi}$ itself. That is also, exactly, the structure of a human -- or another AI system -- trying to keep an increasingly autonomous agent within bounds that were never fully specified in advance, the setting in which agentic misalignment is reported to occur and in which calls for a technical ``brake pedal'' have so far stayed a metaphor. For this reason, we need a full-stack control across all four axes, as this is the safer way to guarantee that the behavior of an AI system stays within the expected range. While the interaction-axis controller of~\eqref{eq:controlled-interaction} is a formal model of that brake pedal, it should be the last line of defense rather than the only one: by the time an intervention reaches this axis, the inner loops have already run to completion.


\medskip
Given the growing interest in safe AI, and particularly in AI control where humanity is shielded from AI systems behaving in undesirable ways, we believe that the Controllable AI framework will be a valuable one. We highlight that, as presented here, Controllable AI is far from being a finished theory, or the only lens through which to view these systems; it is one useful vocabulary among several, and its value will be judged by how much of the above it actually helps close. What we hope this article offers, together with its accompanying open-source implementation, is a shared starting point: a common language in which a skip connection, a steering vector, a scheduled-sampling curriculum, and a reachability certificate are not four unrelated ideas, but four points on the same design space, still waiting for most of the same tools. As Wilbur Wright once said: \emph{``When this [inability to balance and steer] has been worked out, the age of flying machines will have arrived."} We believe that the same will be true about AI.

\bibliographystyle{ieeetr}
    \bibliography{main}

\end{document}